\documentclass[10pt,letterpaper]{article}
\newif\ifpreprint
\preprinttrue 
\newif\ifhilited
\hilitedtrue

\usepackage[top=0.85in,left=2.75in,footskip=0.75in]{geometry}

\usepackage{amsmath,amssymb}

\usepackage{changepage}

\usepackage{textcomp,marvosym} 

\usepackage{cite}

\usepackage{nameref,hyperref}

\usepackage[right]{lineno}

\usepackage[nopatch=eqnum]{microtype}
\DisableLigatures[f]{encoding = *, family = * }

\usepackage[table]{xcolor}

\usepackage{array}

\newcolumntype{+}{!{\vrule width 2pt}}

\newlength\savedwidth

\raggedright
\usepackage[aboveskip=1pt,labelfont=bf,labelsep=period,justification=raggedright,singlelinecheck=off]{caption}

\makeatletter
\renewcommand{\@biblabel}[1]{\quad#1.}
\makeatother

\usepackage{lastpage,fancyhdr,graphicx}
\usepackage{epstopdf}
\fancyheadoffset[L]{2.25in}
\fancyfootoffset[L]{2.25in}
\usepackage{algorithm}
\usepackage{algpseudocode}
\usepackage{calc}                       
\usepackage{siunitx}
\usepackage{versions}
\usepackage{xspace}
 
\newcommand{\figlabel}[1]{\label{fig:#1}}

\newcommand{\fig}[1]{Fig~\ref{fig:#1}}

\newcommand{\Fig}[1]{Fig~\ref{fig:#1}}

\ifpreprint
\newcommand{\Figure}[4]{
  \begin{figure}[#2]
    \centering \includegraphics{#1.pdf}
    \caption[]{\textbf{#3} #4}
    \figlabel{#1}
  \end{figure}
}
\else
\newcommand{\Figure}[4] {
  \begin{figure}[!h]
    \caption[]{\textbf{#3} #4}
    \figlabel{#1}
  \end{figure}
}
\fi

\newcommand{\eqlabel}[1]{\label{eq:#1}}
\newcommand{\eq}[1]{\eqref{eq:#1}}

\newcommand{\seclabel}[1]{\label{sec:#1}}
\newcommand{\secref}[1]{\ref{sec:#1}}
\newcommand{\secn}[1]{Section~\secref{#1}}
\newcommand{\subsecn}[1]{Subsection~\secref{#1}}

\renewcommand{\@}{\partial}             
\newcommand{\abs}[1]{\left\lvert#1\right\rvert} 
\newcommand{\appr}{\sim}
\newcommand{\bydef}{:=}                 
\newcommand{\const}{\mathrm{const}}
\renewcommand{\d}{\mathrm{d}}                       
\newcommand{\df}[2]{\displaystyle{\frac{\partial{#1}}{\partial{#2}}}}         
\newcommand{\dff}[3]{\displaystyle{\frac{\partial^2{#1}}{\partial{#2}\partial{#3}}}}  
\newcommand{\ee}[1]{\mathrm{e}^{#1}}                
\newcommand{\mean}[2]{\overline{\left(#1,#2\right)}}
\newcommand{\Mx}[1]{\begin{pmatrix}#1\end{pmatrix}} 
\newcommand{\mx}[1]{\boldsymbol{#1}}        
\renewcommand{\O}[1]{O\left(#1\right)}				
\newcommand{\Real}{\mathbb{R}}            
\newcommand{\side}[2]{#1_{(#2)}}              
\newcommand{\Zahlen}{\mathbb{Z}}        

\newcommand{\OC}{openCARP\xspace}
\newcommand{\BB}{BeatBox\xspace}

\newcommand{\mm}{\unit{\milli\meter}}
\newcommand{\ms}{\unit{\milli\second}}
\newcommand{\mV}{\unit{\milli\volt}}

\def\p{{\scalebox{0.6}{\text{+}}}}
\def\m{{\scalebox{0.6}{$-$}}}
\def\o{\circ}

\newcommand{\geomname}[2]{\expandafter\def\csname geom:#1\endcsname{#2}}
\DeclareRobustCommand{\geom}[1]{%
  \ifcsname geom:#1\endcsname\csname geom:#1\endcsname
  \else\GenericWarning{}{Warning: geometry `#1' has no \string\geomname}#1\fi}

\geomname{CX122-3}{G1}
\geomname{CT110-3}{G2}
\geomname{CZ110-3}{G2m}
\geomname{CT115-3}{G3}
\geomname{CC115-3}{G3m}
\geomname{CT110-4}{G4}

\newcommand{\+}[2]{\def#1{{#2}}}
\newcommand{\1}[2]{\def#1##1{{#2}}}

\+{\Amax}{A} 
\+{\Afrac}{\mu} 
\+{\aveps}{\epsilon} 
\+{\avk}{k} 
\+{\avlam}{\lambda} 
\+{\avm}{m} 
\+{\a}{a} 
\+{\anode}{A} 
\+{\bslope}{\theta} 
\1{\Bes}{J_{#1}}  
\+{\besind}{\nu} 
\1{\besroot}{q_{#1}} 
\+{\b}{b} 
\+{\bnode}{B} 
\+{\C}{C}  
\+{\Cm}{C_\mathrm{m}}  
\+{\Dhat}{\hat{D}}  
\+{\D}{D}  
\+{\DL}{D_{\parallel}}  
\+{\DT}{D_{\perp}}  
\+{\G}{\mx{G}} 
\+{\F}{F} 
\+{\f}{f} 
\+{\fibang}{\alpha} 
\+{\g}{\mx{g}} 
\+{\gKi}{g_{K1}} 
\+{\gf}{g_f} 
\+{\h}{h}  
\+{\ts}{h_t}  
\+{\ss}{h_x}  
\+{\I}{I} 
\+{\Iion}{I_{\mathrm{ion}}} 
\+{\i}{i} 
\+{\j}{j} 
\+{\K}{K} 
\+{\k}{k} 
\+{\l}{\ell} 
\+{\Lx}{L} 
\+{\M}{M} 
\+{\N}{N} 
\1{\Norm}{\varepsilon_{#1}} 
\+{\tissue}{m} 
\+{\n}{\vec{n}} 
\+{\errn}{n} 
\+{\Domain}{\Omega}
\1{\Domainj}{\Domain_{#1}}
\+{\dDomain}{\@\Domain}
\1{\dDomainj}{\@\Domain_{#1}}
\+{\dDomaini}{\@\Domain_*}
\+{\R}{r} 
\+{\Ri}{r_i} 
\+{\Ro}{r_o} 
\+{\r}{\vec{x}} 
\+{\rp}{\vec{x}^{\,\prime}} 
\+{\sig}{\sigma} 
\+{\sighat}{\hat{\sigma}} 
\+{\sigL}{\sigma_{\parallel}} 
\+{\sigT}{\sigma_{\perp}} 
\+{\svr}{\beta} 
\+{\sS}{\Sigma} 
\+{\t}{t} 
\+{\Tmax}{T} 
\+{\V}{V}  
\+{\Vnum}{V^{\sharp}}  
\+{\Vmin}{\ensuremath{V_{\min}}\xspace}
\+{\Vmax}{\ensuremath{V_{\max}}\xspace}
\+{\Vmid}{\ensuremath{V_{\mid}}\xspace}
\+{\width}{W}  
\+{\w}{w}  
\+{\wp}{p} 
\+{\wq}{q} 
\1{\X}{\ensuremath{\times #1}} 
\+{\x}{x} 
\+{\y}{y} 
\+{\polang}{\phi} 
\+{\polrad}{\rho} 

\+{\jone}{\kappa}
\+{\aone}{a}

\begin{document}
\vspace*{0.2in}
\begin{flushleft}{\Large\textbf{Modelling the electrophysiological interactions
      between human pluripotent
    cell-derived cardiomyocyte grafts and host  ventricular tissue }}
\newline
\\
Suran Galappaththige\textsuperscript{1},
Vadim N Biktashev\textsuperscript{1*},
Faisal J Alibhai\textsuperscript{2},
Michael A Laflamme\textsuperscript{2,3}
\\
\bigskip
\textbf{1} Department of Mathematics, University of Exeter, Exeter, United Kingdom
\\
\textbf{2} McEwen Stem Cell Institute, University Health Network, Toronto, Canada
\\
\textbf{3} Department of Laboratory Medicine \& Pathobiology, University of Toronto, Toronto, Canada
\bigskip

%
%


* \url{v.n.biktashev@exeter.ac.uk}

\end{flushleft}


%
\section*{Abstract}

Human pluripotent stem cell-derived cardiomyocytes (hPSC-CMs) are a promising
therapy for regenerating myocardium after infarction, but their use is limited
by graft-related arrhythmias that frequently occur shortly after
transplantation. Experimental studies indicate that these arrhythmias can
originate within the graft, which may act as an ectopic pacemaker, yet the
mechanisms governing successful excitation of host tissue remain poorly
understood. In particular, the role of electrical coupling at the graft–host
interface is important, but difficult to measure directly or control. Computer
modelling can help here.

Here, we present a computational framework that enables systematic
investigation of graft–host electrical interactions using a
physiologically interpretable parameterisation. We model the
graft–host interface as an internal boundary with a defined specific
conductance, allowing direct control over coupling strength in units
that correspond to measurable tissue properties.
We formulate the governing equations and implement the computations using both
finite-difference and finite-element discretisations in established cardiac
modelling platforms. Using representative anatomical and physiological
configurations, we demonstrate how variations in interface conductance
influence the ability of spontaneous graft activity to initiate propagating
excitation in host tissue.

This framework provides a reproducible, mechanistically transparent tool for
studying graft-related arrhythmogenesis and lays a foundation for evaluating
strategies to mitigate arrhythmic risk in cardiac cell therapy.


%

\section*{Author summary}

One promising treatment to restore damaged cardiac muscle is to replace lost
or supplement existing cells using lab-grown cardiomyocytes derived from human
pluripotent stem cells. A major obstacle is that these implanted cells can
trigger dangerous heart rhythm disturbances shortly after
transplantation. These arrhythmias are thought to arise when the implanted
cells, which can beat spontaneously, electrically stimulate the surrounding
heart tissue and disrupt the normal rhythm.

Experimental studies of when and how this happens can be assisted by
computational modelling of the graft-host interaction. One factor 
that may predict whether a graft is arrhythmogenic is the strength of the
electrical connection between the implanted cells and
the host tissue. Modelling approaches for this do exist, but
 they model this connection using
a random-number generator and are difficult to relate to measurable
biological properties.

In this work, we introduce a new computational method that represents the
connection between implanted and host cells using a physically meaningful
parameter describing electrical conductance. This allows us to systematically
study how coupling strength influences the likelihood that implanted cells
trigger abnormal electrical activity. Our approach provides a clear,
reproducible method for investigating graft-related arrhythmias
and may help guide
strategies to make cardiac cell therapies safer.

\clearpage
\newgeometry{top=0.85in,left=1in,right=1in,footskip=0.75in}
\ifpreprint\else\linenumbers\fi

\section{Introduction}

Human pluripotent stem cell-derived cardiomyocytes (hPSC-CMs) hold
significant promise for regenerating the heart after myocardial infarction
(MI), but their therapeutic potential is limited by graft-related arrhythmias
that arise in the first weeks after cell
transplantation.~\cite{39045008,31056479,29969440,24776797,39196193}
Electro-anatomical mapping of transplanted animals
has demonstrated that the activation site responsible for these
arrhythmias originates from integrated graft tissue, suggesting the
implanted cells function as an ectopic pacemaker. In support of this
concept, gene-editing to eliminate cell automaticity~\cite{37028405}
or direct ablation of initiation sites~\cite{39196193}
reduces the incidence of graft-related 
arrhythmias. Despite this progress, our understanding of the
mechanisms underlying the initiation of these arrhythmias by individual grafts 
remains incomplete. Impulse propagation from the graft to
host tissue depends on numerous factors, including the spontaneous beat
rate of implanted cells, the extent of graft coupling with host
tissue, graft location within the infarcted myocardium, and local
cell-cell interactions.

Here, in silico modelling offers a powerful framework
for systematically dissecting the contributions of different factors
underlying graft-related arrhythmias and evaluating
potential
therapeutic strategies. Accurate computational modelling of the
graft-host interaction requires integrating multiple variables: the
ionic models of graft and host cardiomyocytes, graft localization
within the infarcted myocardium, and the anatomy and connectivity of
cells within the tissue, including fibre orientation and electrotonic
coupling between graft and host. Of these, graft-host connectivity is
particularly critical, as it determines whether spontaneous graft
activity can effectively excite surrounding host tissue to propagate
an arrhythmic impulse. Prior work has identified graft-host
connectivity as a key parameter mediating
graft-related
arrhythmias. For instance, Gibbs et al. \cite{Gibbs-etal-2023} modelled the
variability of this connection using a ``discontinuous finite element
method'', whereby a subset of the edges shared by finite elements
belonging to graft and host tissue were disconnected; the choice of
which edges to disconnect was done randomly. A strength of this
approach is that it can be executed using the \OC software with
only minor modifications to the user interface.
Of course, stochastic modelling is a perfectly
  valid approach in general, and, in a sense, corresponds to the
  fundamental biophysics of inter-cellular connections in this
  particular case.
  However, randomness can make results more difficult to
  reproduce.
  More importantly, it has certain advantages to work with
  a well defined, physically measurable and controllable quantity
  characterizing the tissue coupling, which has certain physical
  dimensionality, e.g. \unit{\siemens/\meter^2} in SI units, or
  \unit{\meter/\second} after accounting for the factors of specific
  cell membrane capacitance and surface/volume ratio. Even though direct
  measurement of this quantity is technically challenging with current
  experimental techniques, having such physical description makes
  modelling more confident, e.g. shows the way to consistency between
  simulations at different spatial discretizations.

The aim of this paper is to introduce the method of modelling
graft-host interaction as an inner boundary with defined specific
conductance, and describe its numerical implementation.  We start by
outlining the mathematical formulation of the problem in terms of
partial differential equations (PDE), proceed to
approaches to
its numerical approximation, and conclude by 
considering some
applications of this constructed model to anatomically and
physiologically realistic examples. We compare our approach to that of
\cite{Gibbs-etal-2023} as a recent and relevant example, but only in
broad terms, as our aim is the description not only within finite
elements, but also in finite differences techniques. Our
finite-difference computations were done with
\BB~\cite{Antonioletti-etal-2017} and finite-element computations
with \OC~\cite{openCARP-paper}.
Both solvers are open source and available respectively:
\OC from \cite{openCARP-sw}
and
\BB from \cite{BeatBox-sw}. 
The histology images, computational
geometries, simulation scripts and parameter files needed
to reproduce selected results of this paper are deposited on
Zenodo~\cite{MEI-data}.

\section{Materials and methods}
\seclabel{materials-methods}

\subsection{Cell electrophysiology and tissue model outline}

To introduce notations: we used monodomain description of the cardiac tissue,
\begin{align}\eqlabel{monodomain}
  \begin{split}
    \df{\V}{t} &= \frac{1}{\Cm} \big[ -\Iion(\V,\g) + \frac{1}{\svr} \nabla\left( \sighat \nabla\V \right)  \big]
    = \f(\V,\g) + \nabla\left( \Dhat \nabla\V \right), \\
    \df{\g}{t} &= \G(\V,\g), 
  \end{split}
\end{align}%
where $\V$ is the transmembrane voltage,
$\Iion$ is the summary transmembrane ionic current,
$\Cm$ is the specific membrane capacitance,
$\svr$ is the surface-to-volume ratio,
$\g$ is the column-vector of local variables such as ionic concentrations and channel gating variables,
$\G$ is the column describing the dynamics of the local variables,
$\f=-\Iion/\Cm$ is voltage dynamics due to transmembrane currents,
$\sighat$ is the conductivity tensor and
$\Dhat=\sighat/(\svr\Cm)$ is the corresponding diffusivity tensor, with
\[
  \sighat=\Mx{\sig_{\j\k}}
  = \Mx{
    \sigT + (\sigL-\sigT)\cos^2\fibang & (\sigL-\sigT)\cos\fibang\sin\fibang \\
    (\sigL-\sigT)\cos\fibang\sin\fibang & \sigT + (\sigL-\sigT)\sin^2\fibang
  },
\]
where $\fibang$ is the fibre direction with respect to the $\x$-axis,
$\sigL$ is the conductivity along fibres and $\sigT$ is the
conductivity across fibres. 

The basic features of our models were similar to the simulations
presented in \cite{Gibbs-etal-2023}, namely
\begin{itemize}
\item The host tissue was modelled using the Ten Tusscher 2006
  ventricular model.~\cite{TenTusscher-Panfilov-2006}
\item The conductivity of the host tissue was
  $\sigL=0.255\,\unit{\siemens/\meter}$ along fibres and 
  $\sigT=0.0775\,\unit{\siemens/\meter}$ across fibres, unless
  specified otherwise. 

\item The graft was modelled using a modified Kernik
    2019 model~\cite{Kernik-19} modified by setting
  $\gKi=0\,\ms^{-1}$ (instead of standard $1.338\,\ms^{-1}$) and
  $\gf=0.087\,\ms^{-1}$ (instead of the standard value
  $0.0435\,\ms^{-1}$). This is the same modification as used
  by Gibbs et. al.\cite{Gibbs-etal-2023}, where it was motivated by
experimental observations of early afterdepolarizations, diastolic
potentials and expression of certain ionic channels. 
  
\item The conductivity of the graft tissue was isotropic, 
  $\sigL=\sigT=0.0775\,\unit{\siemens/\meter}$.
\item In both tissues, $\Cm=1\,\unit{\micro\farad/\centi\meter^2}$, 
  $\svr=1400\,\unit{\centi\meter^{-1}}$,
  so $1/(\svr\Cm)  \approx  7.14286\cdot10^{-4} \;\unit{\meter^3/\farad}$,
  and correspondingly
  $\DL=\sigL/(\svr\Cm)=0.182143\,\mm^2/\ms$,
  $\DT=\sigT/(\svr\Cm)=0.0553571\,\mm^2/\ms$.
\end{itemize}
In test problems designed to verify numerical convergence of our
schemes, we also used the pure diffusion equation for $\V$ without any
reaction, $\f=\mx{0}$; there the ``voltage'' $\V$, space and time were all dimensionless. 
Both in \BB and \OC, time stepping was done as is traditional in
the field, with the Rush-Larsen scheme for gating variables and explicit
Euler for the rest of the dynamic variables, and standard central
difference for \BB and finite element in \OC for space
discretization. 
At the parameters used, the graft
cells spontaneously oscillate with a period of
$533.1\,\ms$ and $\V(\t)\in[-75.6,24.8]\,\mV$, whereas the host
cells are excitable, with the resting state $\V=-85.8\,\mV$
and the action potential range $\V(\t)\in[-86.2,43.1]\,\mV$.

\subsection{Computational mesh from histology}
\seclabel{hist}

Histology sections of infarcted and engrafted hearts
were obtained from
a previous study
investigating hPSC-CM transplantation following ischemia-reperfusion
injury in a swine model\cite{31056479}. Individual graft islands were
identified in scanned histological sections and traced using the
Aperio ImageScope Software (Leica). Histological features were
identified by staining for sarcomeric myosin heavy chain
(Developmental Studies Hybridoma Bank, MF20), human Ku80 (Cell
Signalling), and scar tissue (aniline blue), as previously described\cite{31056479}.
The spatial resolution of the images, which defined the finest
spatial discretization in computations, was
$\ss\approx0.048\,\mm$; for 
illustration presented here, we also used resolutions decreased up to 5-fold in
each direction, that is, a space step $\ss\approx0.24\,\mm$. 

Conversion of the microscopic images was done using typical
image-processing operations, which we implemented as \BB scripts,
following the ideology laid out in \cite{Krinsky-etal-1991}, with the
following main steps.
%
  Separation and thresholding of colour components: used to get
  separately the masks of the muscle tissue, scar tissue and contours
  of the graft tissue, distinguished solely by their colours. 
   Graft tissue mask: obtained
    using the ``detection of closed contours''
  algorithm, as in Fig. 2 of~\cite{Krinsky-etal-1991}. 
  Smoothing of the tissue masks: this was performed using standard dilation/erosion
  method
  as described in~\cite{Krinsky-etal-1991}.
%
The fibre directions were assigned using a rule-based algorithm
similar to the one described in \cite{Gibbs-etal-2023}. 
  The ``whole tissue mask'', comprising host muscle, scar and
  graft, was obtained as a union of the three masks, followed by
  dilation/erosion, similar to the ``restoration of broken
  contour'' algorithm  of
  \cite{Krinsky-etal-1991},  to fill the gaps and ensure that inner and outer
  voids of the mask are disconnected from each other. 
  Identify the outer and inner boundaries:
  the ``closed circuit algorithm'' was used to separately identify 
  the outer boundary (by starting the propagating waves at the outer
  borders of the image rectangle) and the inner boundary (with the
  starting position at a manually selected point within the area
  surrounded by the muscle).
  The Laplace problem was solved in this domain, with Dirichlet
  boundary conditions set at 0 at the inner boundary and at 1 at the
  outer boundary.
  The gradient of the solution of the Laplace problem was computed,
  and the field of fibre direction was defined as orthogonal to the gradient,
  in the ``whole tissue'' domain. 
  The restriction of this direction field to the host muscle
  domain was used to define anisotropic diffusion in calculations. 

A two-dimensional triangular finite-element mesh for monodomain simulations in
\OC was generated from the \BB ``square-pixel'' grid obtained with the above
procedures.  The square lattice was triangulated by splitting every fully
populated quadrilateral cell into two right-angle triangles along a fixed
diagonal, yielding a conforming, $C^0$-continuous mesh whose spatial
resolution matches that of the source image. The triangles that bordered the excluded
pixels were omitted, so that the boundary of the computational domain followed
the segmentation boundary. Each triangle was assigned a single region label by
considering the majority of its three vertices, and a single fibre orientation
by algebraic averaging of its vertex vectors followed by re-normalisation to
unit length. The procedure was to assign a default in-plane direction for
(rare) degenerate cases in which the averaged vector was too small.
After triangulation, vertices that were not
referenced by any element were removed and element indices were
renumbered. The mesh was finally checked for invalid indices, duplicate nodes,
and degenerate (near-zero-area) elements.

\subsection{Finite-difference approximation of diffusion
    and outer boundaries}
\seclabel{fd-std}

 The details of the finite element approximation used by \OC are well described  in the manual of
that software and in the papers referenced
therein.\cite{openCARP-paper,openCARP-sw} The finite difference
approximation used in \BB broadly follows the standard approach as
described in \cite{Clayton-Panfilov-2008}; however, its application
in the present study requires a discussion of issues related to irregular
domain boundaries, strong anisotropy and
discontinuous diffusivity. We approximate the non-flux
boundary conditions by removing the connections between tissue and
non-tissue nodes~\cite{TenTusscher-Panfilov-2008}, rather than by
more elaborate procedures taking into account the position of the
boundary with respect to the grid
nodes~\cite{Forsythe-Wasow-1960,Samarskii-2001}, which are of limited
use when the domain shape is obtained from experiment. The
discontinuous diffusivity requires a conservative
scheme~\cite{Samarskii-2001}. The strong anisotropy, combined with
ragged boundaries, can cause numerical
instability~\cite{Weickert-1998}; instead of more sophisticated
remedies proposed for
this problem~\cite{Gao-Wu-2013-JCP,Ngo-Huang-2016-CCP,Vogl-etal-2023-CMA}, we
use a simple ``polishing'' of the computational grid, removing the
nodes that cause such instabilities. All these details are essential for
reproducibility of our results, but are not directly relevant to our
main message, so they are given in \nameref{S1_Appendix}.

\subsection{Inner boundary: mathematical formulation}
\seclabel{ib-form}

An earlier study by Gibbs et al.~\cite{Gibbs-etal-2023} modelled reduced connectivity between
graft and host tissues using \OC finite elements implementation, by
cutting some of the connections between the finite elements belonging
to the two different kinds of tissue. To control the degree of the
reduction, the authors eliminated a certain fraction of the connections,
choosing them at random.  This served the authors well and
yielded
valuable results, but it is clear that such a method has
some apriori limitations. The most obvious is that the probabilistic
character of
this approach
makes the results less reproducible. Less obvious but
no less significant is that it is not immediately clear how to
modify the fraction of connections to reproduce the simulations
at different spatial resolutions. In the present work, we explored an
alternative formulation. We consider the boundary between host and
graft tissue to be defined as an inner boundary with a certain
specific conductance, rather than being ``all or nothing'' on a
random basis. By formulating these boundary conditions on the PDE
level, we then get a firm guide on how to obtain its discretization
at various spatial resolutions.

To formulate our description mathematically, consider a domain consisting
of two subdomains
$\Domain = \Domainj1 \, \cup \, \dDomaini \, \cup \, \Domainj2$,
corresponding to the two different tissues, where $\dDomaini$ is the
inner boundary,
\[
  \dDomaini = \overline{\Domainj1} \cap \overline{\Domainj2},
\]
and the outer boundaries are
\[
  \dDomain=\overline{\Domain}\setminus\Domain, 
  \qquad
  \dDomainj\tissue = \dDomain\cap\overline{\Domainj\tissue}, 
  \quad
  \tissue=1,2.
\]
For the extension of a function $\F:\Domainj\j\to\Real$ to $\dDomaini$
we shall use notation for the one-sided limit
\[
  \side{\F}{\tissue}(\r) \bydef \lim\limits_{\rp\to\r,\;\rp\in\Domainj\tissue} \F(\rp), \qquad \r\in\dDomaini.
\]
In these terms, the description of the transmembrane voltage is
\begin{align*}
  \df{\V}{t} &= \f_\tissue(\V,\g) + \nabla\left( \Dhat_\tissue \nabla\V \right), \qquad \r\in\Domainj\tissue, \\
  & \n_\tissue(\r)\cdot  (\Dhat_\tissue (\r)\nabla\V (\r))=0, \qquad \r\in\dDomainj\tissue, \qquad \tissue=1,2, \\
  & \n_{12}(\r) \cdot (\Dhat_1(\r) \nabla\side{\V}{1}(\r)) = \sS (\side{\V}{2}(\r) - \side{\V}{1}(\r)), \quad \r\in\dDomaini, \\
  & \n_{21}(\r) \cdot (\Dhat_2(\r) \nabla\side{\V}{2}(\r)) = \sS (\side{\V}{1}(\r) - \side{\V}{2}(\r)), \quad \r\in\dDomaini, \\
\end{align*}
where $\n_\tissue$ is the outer unit normal to $\dDomainj\tissue$, $\n_{12}$ is the unit normal to $\dDomaini$ from $\Domainj1$ into $\Domainj2$,
and $\n_{21}=-\n_{12}$  is the unit normal to $\dDomaini$ from
$\Domainj2$ into $\Domainj1$, and $\sS$ is the specific electrical
  conductance of the inner boundary, i.e.\ conductance per unit boundary area,
  scaled by membrane capacitance and surface/volume ratio in the same way as
  $\Dhat$, hence measured in $\mm/\ms$; we call it the interface conductance for short.

\subsection{Finite difference approximation of inner boundary}
\seclabel{ib-appr}

Finite difference approximation of this inner boundary was constructed
by modifying the weights connecting the nodes occurring on different
sides of the boundary. That is, we treat near-boundary nodes in the
same loops as the inner nodes, as all calculations pertinent to the
boundary are done at the preliminary stage, at the same time as
constructing the precomputed weights for the stencils describing the
diffusion operator.  The formulas for the modifications were obtained
according to the following principles (again, using the standard language of
electrical circuits):
\begin{itemize}
\item Assuming that the inner boundary is smooth and we are interested
  in the limit of small $\ss$, we consider the boundary to be a
  straight line, of a certain slope, say angle $\bslope$ with respect
  to the $\x$-axis.
\item Consider a piece of the straight boundary of unit length, and
  let the voltage difference between the two sides of this boundary be
  1. Then the total current through this piece of boundary will, by definition,
  be equal to $\sS$.
\item The total current through the boundary will be provided via
  connections crossing it. Each such elementary current numerically
  equals the conductance of the connection, since the voltage drop
  across this connection is 1 by assumption.  The conductance of the
  connection is its weight in the stencils of the two connected nodes
  (remember these weights must be symmetric for charge
  conservation). Hence, $\sS$ is numerically equal to the sum of all
  the weights of connections crossing the unit length of the border.
\item We intend to exploit the same stencils for the inner boundary as
  are used for the diffusion operator, just modifying the weights.
  Hence, there are only a few possible directions which the connections
  may have.  If we use a five-point stencil, then there are two
  possible directions: horizontal, for $\w_{\p\o}$ and $\w_{\m\o}$,
  and vertical, for $\w_{\o\p}$ and $\w_{\o\m}$.  If we use a
  nine-point stencil, then there are in addition two diagonal
  directions, the ``upward diagonal'' for $\w_{\p\p}$ and $\w_{\m\m}$, and
  the ``downward diagonal'' for $\w_{\p\m}$ and $\w_{\m\p}$.
\item Let us count the connections in each of the four directions, 
  and then add up these four contributions to get the total $\sS$. The
  number of connections of a certain direction crossing the unit
  length of the inner boundary can be computed as the size of this
  unit length in the direction orthogonal to the connections,
  divided by the transversal distance between connections, that is the
  distance between parallel lines passing through the
  connections. For a boundary making angle $\bslope$ with the
  $\x$-axis, the size of a unit piece of boundary across the
  horizontal connections is $\abs{\sin\bslope}$, for the vertical
  connections it is $\abs{\cos\bslope}$ and for the diagonal ones is
  $\abs{\sin(\bslope\pm\pi/4)}=\abs{\sin\bslope\pm\cos\bslope}/\sqrt2$. The
  transversal distance between horizonal connections is $\ss$ and the
  same between vertical ones, whereas for the diagonal connections it
  is $\ss/\sqrt2$.
\item As to the relative weights of connections in the four
  directions, in absence of reasons to think of inner boundary as
  anisotropic, it seems to make sense to observe symmetries
  $\x\leftrightarrow-\x$, 
  $\y\leftrightarrow-\y$ and
  $\x\leftrightarrow\y$, but otherwise, 
  the above considerations leave open the question about the
  relative weights of connections in the straight and in the diagonal
  directions. 
  We considered two variants, by modelling these relative weights on two ``standard'' stencils,
  the 5-point stencil, and the 9-point stencil known as
  Patra-Karttunen scheme~\cite{Patra-Karttunen-2006} (although known well
  before 2006, see e.g. ~\cite{Barkley-etal-1990} and references
  therein). That is, we considered
  (see \subsecn{av-appr} for definition of $\w_{\k\l}$)
  \[
    \Mx{ 
      \w_{\m\p}  & \w_{\o\p} & \w_{\p\p} \\
      \w_{\m\o}  & \w_{\o\o} & \w_{\p\o} \\
      \w_{\m\m} & \w_{\o\m} & \w_{\p\m} 
    }=\K\Mx{ 
      \wq  & \wp & \wq \\
      \wp  & *     & \wp \\
      \wq  & \wp & \wq
    }
  \]
  where $\wp=1$, $\wq=0$ for the five-point stencil, and
  $\wp=2/3$, $\wq=1/6$ for the Patra-Karttunen 9-point stencil, and
  the coefficient $\K$ is to be chosen so as to provide the required
  value of $\sS$.
\item As a result, considering that
  $\abs{\a+\b}+\abs{\a-\b}=2\max(\abs{\a},\abs{\b})$,
  we get
  \[
    \K = \frac{\sS / \ss}{
      \wp(\abs{\cos{\bslope}}+\abs{\sin{\bslope}})
      +
      2\wq\max\left(\abs{\cos{\bslope}},\abs{\sin{\bslope}}\right) 
    } ,
  \]
  and the stencil weights corresponding to connection across boundary
  were replaced correspondingly with $\wp\K$ for straight (horizontal
  or vertical) and with $\wq\K$ for diagonal connections. 
\end{itemize}

The final detail required for the implementation of the inner boundary
condition is determination of the local slope of the boundary at each
point where the trans-boundary connection is to be determined. In
accordance with the discussion in \subsecn{nonflux-appr}, in real
life this information is to be obtained from the grid itself. 

We have used the ``dipole moment'' heuristic:
\begin{itemize}
\item assign equal and opposite ``electric charges''
  to the nodes corresponding to the opposite sides of the boundaries;
\item for each of the nodes related by the connection in question,
  determine the dipole moment of the nodes in its $3\times3$ vicinity;
\item calculate the dipole moment of the connection as
  the sum of the dipole moments of the two nodes it connects;
\item consider the direction of this summary dipole moment as the
  local normal to the boundary. 
\end{itemize}
In this case, the nodes that belong to the intersection of the
$3\times3$ vicinities of the two connected nodes get accounted twice;
we did not consider this of any importance as this is a heuristic
anyway.

Finally, we note that this way of determining the orientation of the
boundary can produce only a finite number of answers; this number is
rather big and sufficient for practical purposes, but in principle it
sets a limit for the achievable accuracy when considering numerical
convergence as $\ss\to0$, because this number remains unchanged as the
discretization gets refined, and so the associated error does not
disappear. We do not know an easy way to assess apriori how big this
error is. The only heuristic argument is that a nine-point stencil is
preferable to a five-point one as averaging the error over a
larger
number of connections
yields a larger
number of possible directions so
should reduce the overall effect,
resulting in
a better
approximation of the continuous spectrum of boundary orientations for purely statistical
reasons.

\subsection{Tests for inner boundary approximations}
\seclabel{ib-test}
  The numerical approximation described in \subsecn{ib-appr}
was validated by problems with exact analytical answers.
For 1D, the problem was posed for $\Domain_1=(-\pi/8,0)$, $\Domain_2=(0,\pi/4)$, $\D_1=3/4$,  $\D_2=3$, $\sS=1$,
with an exact solution
\begin{equation} \eqlabel{1j-exact}
  \V(\x,\t)=\ee{-3\t} \times
  \begin{cases}
    \sqrt2 \cos(2\x+\pi/4), \quad \x\in\Domain_1, \\
    -\sqrt2 \cos(\x-\pi/4), \quad \x\in\Domain_2.
  \end{cases}  
\end{equation}

For 2D, the problem was posed as
$\Domain_1\in\{(\x,\y)\,|\,\polrad\in[0,1)\}$, 
$\Domain_2\in\{(\x,\y)\,|\,\polrad\in(1,2)\}$,
$\D_1=1$,
$\D_2=\jone/2$,
$\sS=\aone\Bes1(1)/ ( \aone\Bes0(1) - \Bes0(\jone/2)  )$,
where
$\polrad=\sqrt{\x^2+y^2}$,
$\jone=3.83\dots$ is the first positive root of $\Bes1()$,
$\aone=2\Bes1(\jone/2)/(\jone\Bes1(1) )$,
with an exact solution
\begin{equation}\eqlabel{2j-exact}
  \V(\x,\y,\t) = \ee{-\t} \times
  \begin{cases}
    \aone \Bes0(\polrad), \quad (\x,\y)\in\Domain_1, \\
    \Bes0(\jone\polrad/2), \quad (\x,\y)\in\Domain_2.
\end{cases} 
\end{equation}

\subsection{Border zone as an approximation of inner boundary}
\seclabel{ibz}

The standard user interface of \OC does not allow immediate
implementation of the inner boundary as defined above. Our practical
approach to overcome this obstacle is  to replace an ``infinitely
thin'' inner boundary with a strip $\Domainj3$ of a finite width
$\width$, with no ionic kinetics, i.e. pure passive diffusion in it,
and diffusivity chosen so that it emulates the infinitely thin
boundary:
\begin{align*}
  \df{\V}{t} &= \f_\tissue(\V,\g) + \nabla\left( \Dhat_\tissue \nabla\V \right), \qquad \r\in\Domainj\tissue, \quad \tissue=1,2,3 \\
  & \f_3=\mx{0}, \qquad \Dhat_3 = \Mx{\D_3 \delta_{\k,\l}}, \quad  \D_3=\sS \width,
\end{align*}
and no artificial inner boundaries between $\Domainj1$ and $\Domainj3$
and between $\Domainj3$ and $\Domainj2$, i.e. the transmembrane
voltage $\V$ and its flux $\Dhat\nabla\V$ are presumed continuous
throughout $\Domain$. Notice that here the diffusivity tensor is
a discontinuous function of the space coordinates, making
\subsecn{av-appr} essential.

The approximation achievable by this approach is limited by two
distinct factors. Firstly, in order to approximate the ``infinitely
thin'' boundary, the border zone width $\width$ should be small
compared to typical geometric scales of the problem, i.e. the smallest
relevant anatomical feature. On the other hand, $\width$ must be large
compared to spatial discretization step $\ss$, for the diffusivity
within the border zone to be adequately described by the diffusion PDE;
particularly challenging aspect here is that the above idealized
description is based on the assumption of $\width$ being constant
along the border zone, which is not realistic if $\width\sim\ss$ and
the border zone has a complicated shape (i.e. is not parallel to
either coordinate axis along all its length). Thus, in a realistic
situation when one has to commensurate the required accuracy of the
results with the computational cost, one may expect qualitative and,
at best, very approximate quantitative correspondence between finite
$\width$ with $\width\to0$ results. We believe that this approach
still has practical value, if the inaccuracy introduced by this
simplification is comparable with indeterminacy inherently associated
with limited accuracy of experimental data in all important aspects,
including ionic models, anatomy of the tissues and diffusion tensor
fields.

\subsection{Idealized geometry}
\seclabel{ideal-def}

  To illustrate and test our inner-boundary model, we used an intermediate level
  of realism, with a realistic ionic
  model for graft and host tissues and ``gradual'' description of the
  boundary conductance as per the theory laid out in \subsecn{ib-appr}, 
  but still an idealized shape of the border between them.
  In this series of simulations, the border was of circular shape of radius $\R$ with the
  graft tissue being within the circle and the host tissue outside of it,
  or vice versa. We used
  isotropic diffusivities, smaller $\D=\DT$ for the graft and larger,
  $\D=\DL$ for the host. Exploiting the symmetry of the problem and to
  make calculations more efficient, we limited them to the first
  quadrant, that is considering quarter-disk of radius
  $\R$ rather than
  whole disk: $\Domain_1=\{(\x,\y)\,|\,\x^2+\y^2<\R^2,\;\x>0,\;\y>0\}$.
  Since on this occasion we do not have an exact analytical solution to
  compare to, we used an overall domain of a square shape, $\Domain=(0,\Lx)^2$,
  with $\Lx>\R$. In simulations with finite-width border
  zone as defined in \subsecn{ibz}, we defined
  $\Domain_3=\{(\x,\y)\,|\,\R^2<\x^2+\y^2<(\R+\width)^2,\;\x>0,\;\y>0\}$
  with $\Lx>\R+\width$.
  Correspondingly, the ``outer'' zone $\Domain_2$ was defined simply as the inside
  of $\Domain\setminus\Domain_1$ for the zero-width border and
  of $\Domain\setminus(\Domain_1\cup\Domain_3)$ for finite-width
  border. 
  The finite-width border zone arrangement is illustrated in
  \fig{quad-mesh}, both for \BB and \OC simulations. Panel (b)
  also shows the position of the point in $\Domain_2$ used for checking
  if the host tissue was excited during the simulation; in \BB
  simulation it was done in a similar way. 
  
\Figure{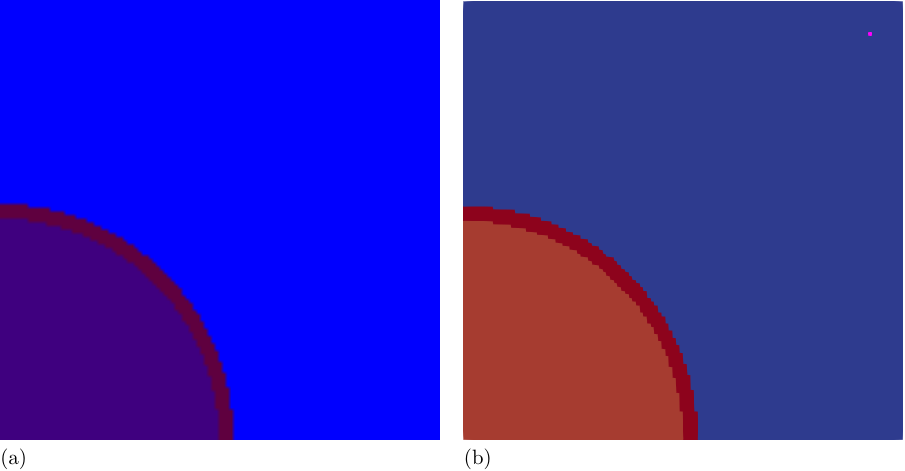}{!htb}{An illustration of the quarter-disk.}{%
  A quarter-disk of radius
   $\R=3\,\mm$  of graft tissue with border
  zone within host tissue box size $\Lx\times\Lx=6\,\mm\times6\,\mm$, border zone width
  $\width=0.2\,\mm$, spatial discretization
  $\ss=0.05\,\mm$. %
  (a) \BB, colour coding
  representing distribution of the diffusivity,  red for low, blue
  for high not to scale. %
  (b) \OC, standard visualization with colours representing tissue tags: dark red for border
  zone $\Domain_3$, lighter red for graft $\Domain_1$, blue for host
  $\Domain_2$. %
  The pink marker in (b) indicates the distal host observation point
  used to register excitation of the host tissue.
}

\section{Results}
\seclabel{results}

\subsection{Construction of the computational mesh}
\seclabel{mesh-def}

\Figure{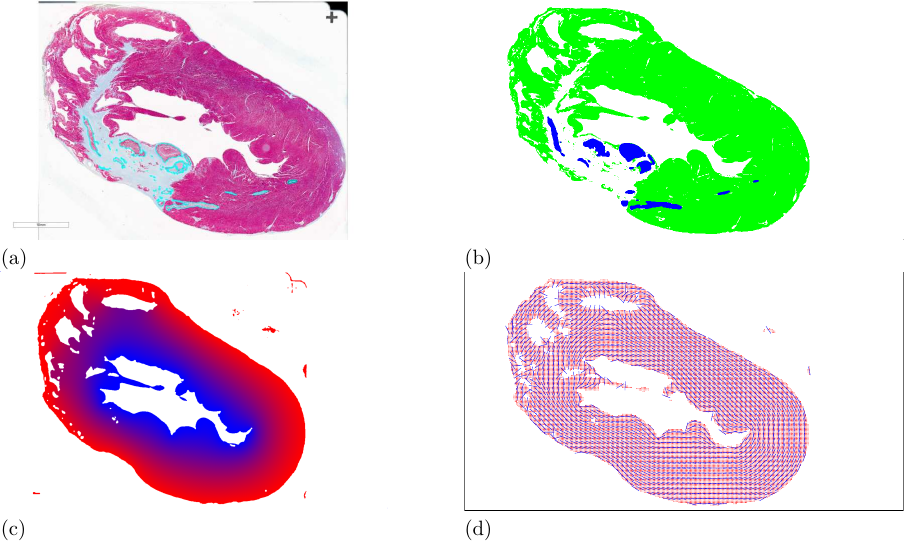}{!htb}{Construction of the computational mesh.}{%
    (a) A histological image (pink: muscle tissue, pale blue:
    scar tissue) with graft position drawn by
    hand (cyan). %
    (b) The resulting computational grid, 
    green pixels are host tissue,
    blue pixels are graft tissue. %
    (c) Solution of the Laplace equation within the whole tissue
    (muscle+scar). %
    (d) Fibre directions in the whole tissue as normals to the
    gradient of the solution to the Laplace equation.
}
  
\Fig{tpef} illustrates the result of the procedure of building the
computational mesh, described in \subsecn{hist}, for one of the
histology images, referred to as \geom{CT110-3}.  We also consider
coarse-grained versions of the resulting meshes, at the spatial
resolutions of $\ss\approx0.04781\times\M\,\mm$, with $\M=1,2,3,5$,
which we shall refer to as \X1, \dots, \X5 for brevity, so \X1
corresponds to the original ``fully resolved'' grid.

\Figure{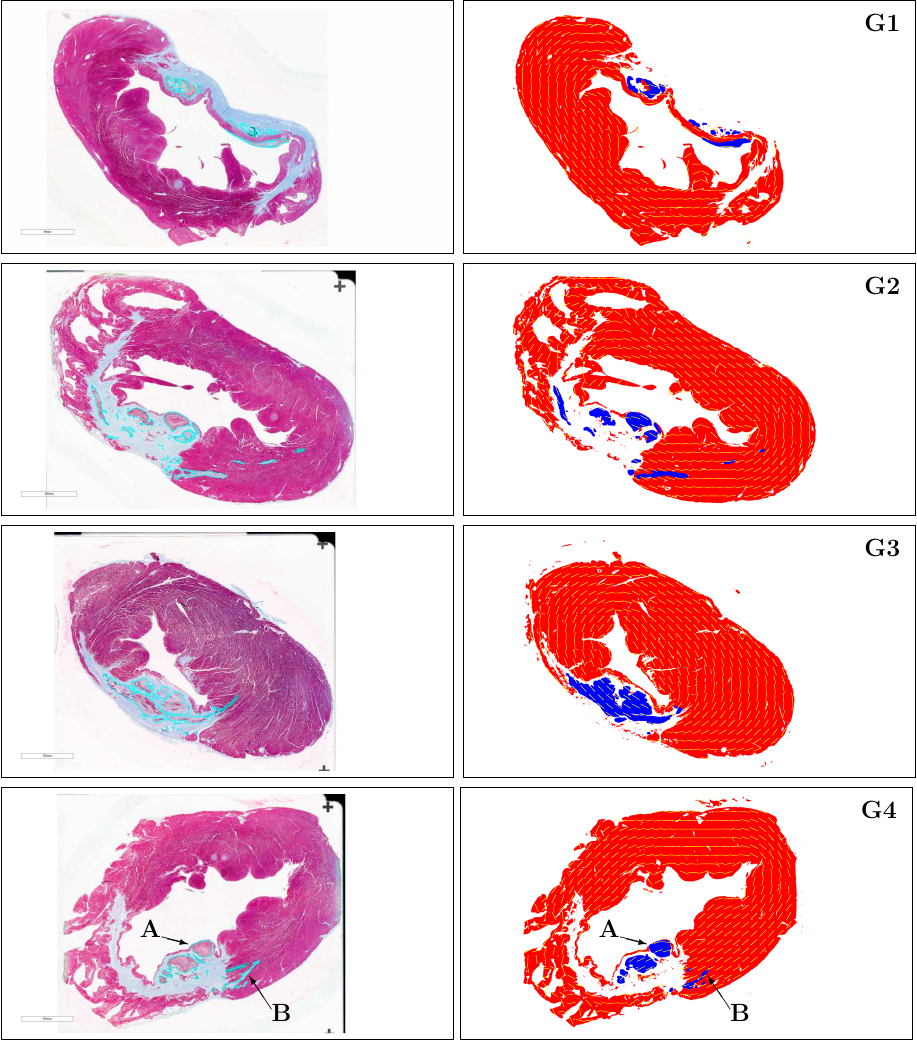}{!htb}{The four selected histology images}{%
  For each geometry we show the original image with grafts labelled on
  the left, and the resulting \BB geometry on the right, with host
  tissue in red, graft tissue in blue and fibre direction with yellow
  strokes.
  In the \geom{CT110-4}, labels `A' and `B' point at the two
  disconnected graft patches that play different roles as explained later.
}

  \Fig{allgeom} introduces the four histology images and corresponding
  geometries, named
  \geom{CX122-3},
  \geom{CT110-3},
  \geom{CT115-3},
  \geom{CT110-4},  
  which we have selected for our simulations presented
  below. Two of them were used with some variations which we explain
  in due course; these variations are visually minor and so not
  presented in \fig{allgeom} to reduce clutter. 

\subsection{Testing of numerical approximations}

\subsubsection{Inhomogeneous diffusion}

\Fig{2avchb-rev} illustrates 
the analytical solution to the test problem with inhomogeneous
anisotropic diffusion, described in \subsecn{av-test},
and numerical convergence to it as a function of
$\ss$, with $\ts=0.12\,\ss^2$, in terms of three error norms,
  \begin{equation}\eqlabel{3norms}
    \begin{split}
      \Norm{\errn} & \bydef \left[ \frac{
          \int_0^{\Tmax}\int_{\Domain} \abs{ \V(\x,\y,\t) - \Vnum(\x,\y,\t) }^{\errn}\,\d\x\d\y\d\t
        }{
          \int_0^{\Tmax}\int_{\Domain} \abs{ \V(\x,\y,\t) }^{\errn}\,\d\x\d\y\d\t
        }
      \right]^{1/\errn},
      \qquad
      \errn=1,2,
      \\[2ex]
      \Norm{\infty} & \bydef \frac{
        \max\limits_{\t\in[0,\Tmax]} \max\limits_{(\x,\y)\in\Domain} \abs{ \V(\x,\y,\t) - \Vnum(\x,\y,\t) } 
      }{
        \max\limits_{\t\in[0,\Tmax]} \max\limits_{(\x,\y)\in\Domain} \abs{ \V(\x,\y,\t) } 
      },
    \end{split}
  \end{equation}
where $\Vnum$ stands for the numerical solution,
all integrals evaluated by the trapezoidal rule,
and $\Tmax$ is total simulation time. Here we took
$\Tmax=0.02$, so $\exp(-\avlam\Tmax)\approx0.187$. 

\Figure{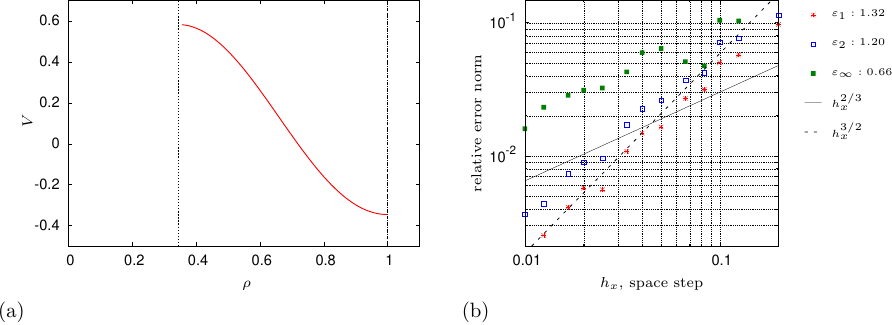}{!htb}{Test of approximation of the anisotropic spatially variable diffusion.}{%
  (a) The $\polrad$-profile of the analytical solution
  \eq{avsol} (\subsecn{av-test}) at $\t=0$, $\polang=0$. %
  (b) The three norms of the
  approximation error as functions of space discretization step.
}

\subsubsection{Suppression of anisotropy-caused instability}

\Figure{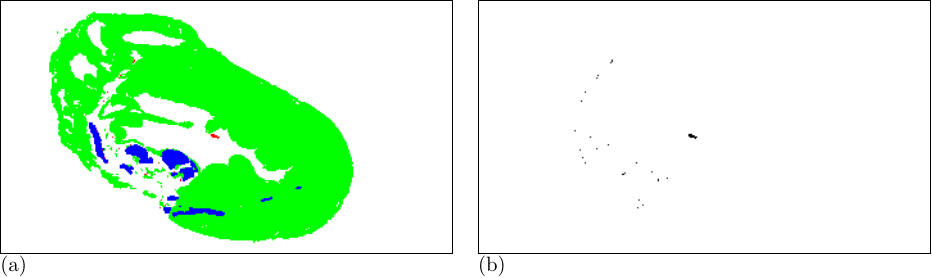}{!htb}{Work of the ``polishing'' algorithm}{%
  on \geom{CT110-3} at \X5 resolution. 
  (a) Computational grid (blue and green as in \fig{tpef})
  relieved from instability-causing points (red). 
  (b) Instability-causing points separately (black). 
}

As discussed in \subsecn{polishing}, 
sufficiently strong anisotropic diffusion operator in domains with
rough boundaries can lead to numerical instabilities that do not
depend on time step. There we also present a ``grid polishing''
Algorithm~\ref{alg:polishing} to tackle this issue. 
A result of this algorithm is 
illustrated in \fig{polished-rev} on the same histology geometry as in
\fig{tpef}.
This is done on the \X5 reduced version
($\ss\approx0.24\,\mm$).  In this example,
40 ``instability causing'' points were removed  out
of 22111, that is about $0.18\%$ of the total.
Over all the geometries and all resolutions subjected to this
polishing algorithm, the average fraction of removed points was
$0.14\%$ and median $0.074\%$.

\subsubsection{Inner boundary approximation results}

The convergence of the numerical solution to the exact solution~\eq{1j-exact}
is illustrated by \fig{1j-rev}, in terms of the same three  norms \eq{3norms},
\textit{mutatis mutandis} (in particular, without integration by $\d\y$).
 The points are
aligned along three distinctive lines for any of the three series,
$\Norm1$, $\Norm2$ and $\Norm\infty$.  The explanation
for this is
related to the point made in \subsecn{nonflux-appr} about
the dependence of the boundary condition approximation on the exact
position of the boundary with respect to the grid nodes. The same
argument applies here to the inter-tissue boundary. Namely, the
numerical scheme approximating the inner boundary, as
described in \subsecn{ib-appr}, corresponds to the boundary being exactly halfway between
the marginal nodes of the two tissues.  The setting of the
simulations underlying \fig{1j-rev} was
such that this was true if and only if the number of grid nodes was divisible by 3.  Correspondingly, the three lines correspond to different remainders after division by three; the best approximations are where the number of grid nodes is fully divisible by 3, where the inner boundary is precisely where it is supposed to be by the inner boundary condition discretization formula.

\Figure{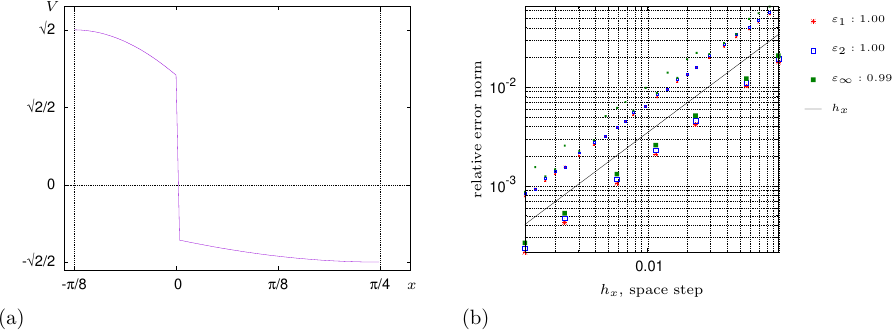}{!htb}{Convergence test in 1D.}{%
  (a) Initial condition for the exact solution. %
  (b) Numerical convergence, log-log plot. %
  Big symbols as defined by the legend correspond to the number of
  points in the grid divisible by 3, as opposed to those represented
  by the corresponding small symbols.
  $\Norm1$ means the average absolute error of the numerical solution vs
  the exact solution,
  $\Norm2$ is the root mean square error, and
  $\Norm{\infty}$ is the maximum error,
  the legend shows the ``best fit'' power index for each error norm for
  the subset of ``big symbols''. 
  The line $\const\times \ss$ is added to guide the eye. 
}

The convergence of the numerical solution to the exact solution~\eq{2j-exact}
is illustrated by \fig{2j-rev}.  The $\O{\ss}$ approximation is roughly
observed in the shown range of $\ss$, worse at smaller ones; we
attribute it to the limit associated with the empirical definition of
the slope of the inner boundary, discussed at the end of \subsecn{ib-appr}.

\Figure{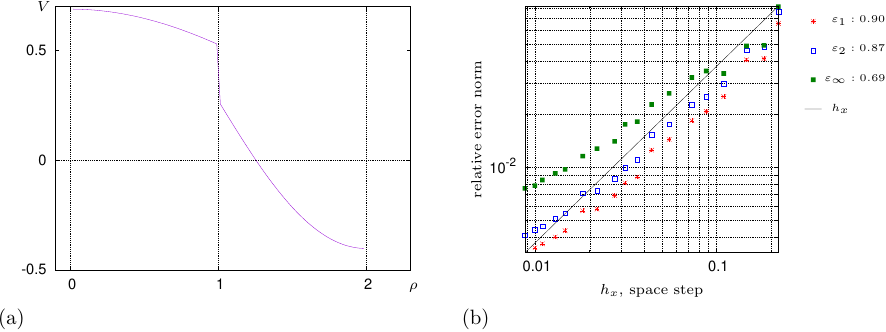}{!htb}{Convergence test in 2D.}{%
  (a) Radial profile of the initial condition for the exact solution
  ($\t=0$, $\polang=0$).  %
  (b) Numerical convergence, log-log plot, notations the same as in \fig{1j-rev}.
}
\subsection{Graft-host interaction with random  decoupling}
\seclabel{stoch}

\Figure{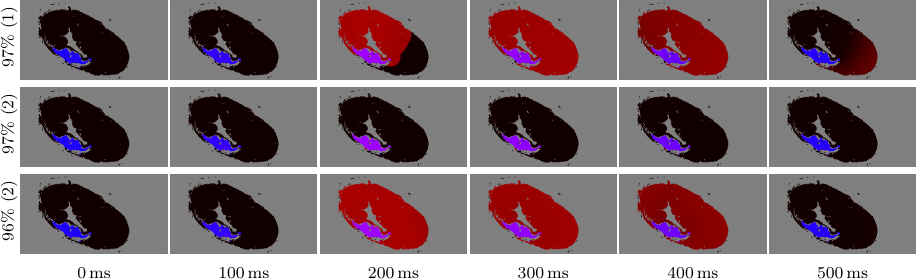}{!htb}{Effects of random decoupling.}{%
  Upper row: 97\% suppression is not
    sufficient to block propagation with random number generator seed 1. %
    Second row: the same 97\% suppression  with random seed 2 does
    block propagation. %
    Bottom row: same random seed 2 but the 96\% suppression does allow
    propagation.  
    Here and below, the colour coding is: the red
    colour component represents $\V$ (excitation), the blue component
    labels the graft tissue. This is for \geom{CT115-3} at \X5-fold reduced resolution
    ($\ss\approx0.24\,\mm$). 
  }

  We have roughly reproduced the model of decoupling the graft from
  host, described in \cite{Gibbs-etal-2023}, namely, cutting a randomly
  chosen fraction of connections between the neighbouring graft and host
  cells. Three examples illustrating resulting simulations are shown in
  \fig{stoch-rev}. We have scanned the fraction of suppressed (cut)
  connections with the step of one percent, i.e. 0.01.  The
  expectation is that when the
  decoupling fraction is below a threshold, oscillations in the graft
  penetrate the host tissue and cause it to oscillate as well, and above
  that threshold, the oscillations remain localized in the graft tissue,
  whereas the host remains in the resting state (there are no sinus
  rhythm sources in our simulations).  Since the procedure is inherently
  probabilistic, we have done that with different values of random number
  generator seed. As a result, the same percentage of cut connection may
  or may not ensure suppression of propagation depending on chance, so
  the above mentioned threshold is not well defined.
  
  \Figure{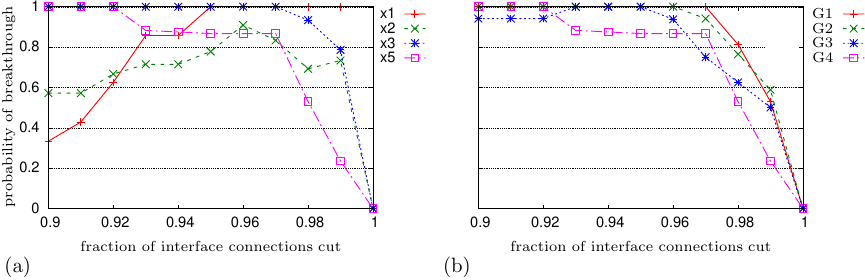}{!htb}{Probability of stochastic decoupling}{
    as a function of the fraction of suppressed connections,
    (a) for
    \geom{CT110-4} at various resolutions as indicated in the legend;
    (b) for resolution \X5, in various geometries as indicated in the
    legend. The vertical axis is common for both panels.
  }
  
  To give a more representative picture in view of the random nature
  of this description, we have repeated simulations, for each
  combination of geometry, resolution and fraction of cut connections
  with $\N=17$ different seeds of the random number generator, from 0
  to 16 inclusive. More precisely, since we applied the polishing
  algorithm to the geometries before randomly cutting the connections,
  this sometimes led to numerically unstable grids and simulations did
  not complete, so in some cases we had $\N<17$.  \Fig{dpp-graphs}
  illustrate the probability of suppression of the host ignition for
  such samples, as a function of resolution for one of the geometries,
  and the function of geometry for the crudest resolution. What we see
  there is that typically an overwhelming fraction of connections
  needs to be cut out to ensure block of propagation of excitation
  from graft to host, that this is very dependent on chance due to
  inherent randomness of the procedure, and that behaviour at
  different resolutions is far from predictable and may be outright
  paradoxical, see the \X1 and \X2 curves in panel (a).

  Of course, proper Monte Carlo style simulations would require a much
  larger sample sizes $\N$, and if we were to polish each geometry
  after randomly cutting connection through interface, this would
  further hugely increase the amount of calculations, so curves
  presented in \fig{dpp-graphs} should not be considered as results
  \textit{per se}, but merely as an illustration of what sort of
  difficulties one is likely to face if using this approach in
  finite-difference setting.

\subsection{Graft-host interaction via inner boundary: idealized geometry}
\seclabel{ideal-result}

\Fig{RS-portrait-rev} summarises the results of two-parametric studies
with idealized geometry as defined in \subsecn{ideal-def},
graft tissue in $\Domain_1$ and host tissue in $\Domain_2$, with
varying $\R$ and $\sS$, at $\Lx=6\,\mm$. Slightly oversimplifying, in these
simulations, there were three typical scenarios: %
1)
  The host tissue suppresses the spontaneous oscillations of the
  graft, and the whole preparation comes to a stationary state after a
  quick transient depending on the details of the initial conditions
  (red crosses in \fig{RS-portrait-rev}). 
2)
  The spontaneous oscillations happen in the graft but are isolated,
  i.e. do not engage the host tissue
  (blue open circles in \fig{RS-portrait-rev}). 
3)
  The spontaneous oscillations evoke an excitation response of the
    host tissue
  (green-filled circles in \fig{RS-portrait-rev}).     

\Figure{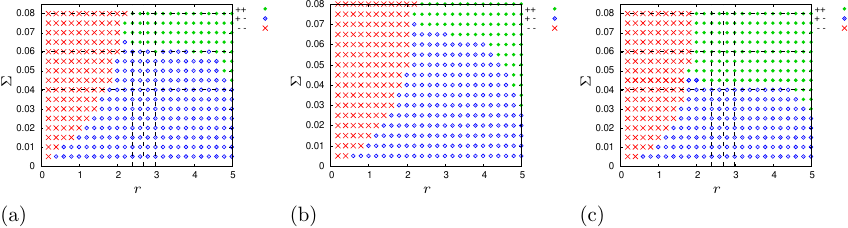}{!htb}{%
  Graft boundary radius
  $\R$
  vs interface conductance
  $\sS$
  parametric plots.
}{
  (a) Finite differences (\BB), zero-width inner boundary ($\width=0$); %
  (b) Finite differences (\BB), border zone emulation ($\width=0.2\,\mm$); %
  (c) Finite elements (\OC), border zone emulation ($\width=0.2\,\mm$). %
  Green filled circles: oscillations registered both in the graft and
  in the host (`$++$'). %
  Blue open circles: oscillations registered in the graft but not in
  the host (`$+-$'). %
  Red crosses: oscillations not registered neither in the graft nor in
  the host (`$--$'). 
  The black dashed lines in (a) and (c) correspond to the
  cross-sections used for \fig{RS-periods-rev} below. 
}

For the purposes of this study, we did not make any finer
distinction between different scenarios: the question of whether
oscillations were observed in $\Domain_1$ and in $\Domain_2$ were
decided simply based on the analysis of the amplitude of the voltage,
\(
\Delta\V \bydef 
\max\limits_{\t\in[0,\Tmax]} \V(\x_\j,\y_\j,\t)
-
\min\limits_{\t\in[0,\Tmax]}\V(\x_\j,\y_\j,\t)
\),
measured at the selected points $(\x_\j,\y_\j)$, $\j=1,2$, where
$(\x_1,\y_1)$ was at the grid node nearest to the centre of the disk,
i.e. bottom left corner in \fig{quad-mesh}, guaranteed to be in $\Domainj1$, and
$(\x_2,\y_2)$ in the top right corner, i.e. a point guaranteed to be
in $\Domainj2$. This crude classification ignores borderline cases,
e.g. Wenckebach-style frequency division when the host responds to some graft
oscillations, but not to each of them. 
Given the period of spontaneous oscillations of the
graft, we found that $\Tmax=2000\,\ms$ was
sufficient.
In the \OC simulations, which were run for
$\Tmax=6000\,\ms$, oscillations were instead detected by counting
beats, i.e. upstrokes of $\V$ through $-40\,\mV$, at mesh nodes near
the same two positions.

The similarity between the panels in \fig{RS-portrait-rev} illustrates our
point that the finite-width border zone description is a qualitatively
correct approximation of the zero-width one, albeit rather crude
quantitatively. All three parametric portraits look similar, in
particular
with
isolated graft oscillations occurring when interface conductance $\sS$ is small,
and graft oscillations suppressed when curvature radius
$\R$ is
small, so  the graft lacked sufficient ``source'' current to overcome
the massive electrotonic load (sink) of the host tissue, even at high
interface conductance.

Note that the propagation/no propagation (green/blue) boundary
decreases to lower $\sS$ near the right end of the diagram; this is
because the maximal value  $\R=5\,\mm$  is rather close
to the box size $\Lx=6\,\mm$, so the boundary effect
is significant: the right $\x=\Lx$ and the upper $\y=\Lx$ non-flux
boundaries help excitation of the host.

Apart from the yes/no (oscillation/no oscillation) answer, of
considerable interest for practice is the period of the oscillations,
as other things being equal, faster graft oscillations are more
arrhythmogenic. \Fig{RS-periods-rev} illustrates the dependence of the
period of graft and host oscillations on the control parameters $\R$
and $\sS$. The plots show both the periods of the host (bigger
symbols) and graft (smaller symbols). We note that when both are
present, they are equal, that is, Wenckebach or other
borderline phenomena, even if they took place, are not present in the
selection. Another evident feature is apparent discontinuity, or at
least a rather quick change, of the
period of graft oscillations across the propagation/no propagation
boundary. 
One important observation is the significant slowing
of oscillations at smaller graft sizes. The available data are
consistent with the scenario that the period tends to infinity as the
radius approaches a certain critical radius depending on $\sS$, 
which is suggestive as to the possible mechanisms of this critical
behaviour from the dynamical systems viewpoint (which goes beyond the
scope of the present study). From the application viewpoint, this is
notable as it means that smaller size grafts are likely to be less
arrhythmogenic.

\Figure{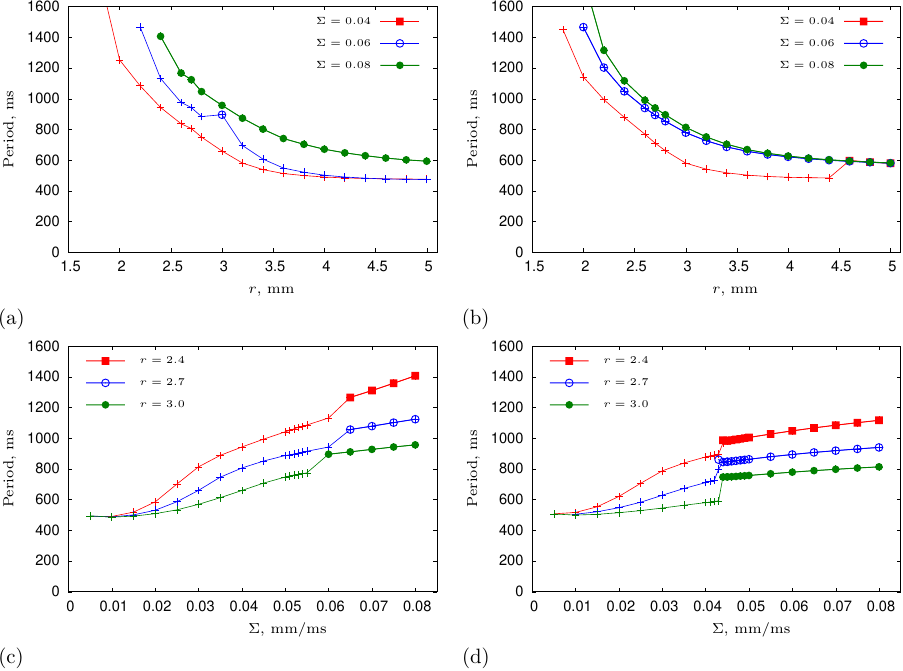}{!htb}{Period of oscillations.}{
  (a,b) as a function of graft 
  radius, at selected values of $\sS$ (in $\mm/\ms$),
  and %
  (c,d) as a function of interface conductance $\sS$, at
  selected values of $\R$ (in $\mm$).  %
  (a,c) Finite differences (\BB), zero-width inner boundary; %
  (b,d) Finite elements (\OC), border zone emulation. %
  Large symbols indicate host oscillation periods;
  small symbols indicate graft oscillation periods. 
}

\subsection{Graft-host interaction via inner boundary: realistic geometry}
\seclabel{real}

 Finally, we apply the
  inner-boundary description to the ``realistic'' geometries, that is
  those obtained from histology images and with rule-based anisotropy,
  as explained in \subsecn{mesh-def}.  Here we consider the selected
  geometries described above, at the spatial resolutions 
  \X1, \X2, \X3 and \X5.  Since the geometries are
  from a fixed set, there was no continuous parameter corresponding to
  $\R$ from \subsecn{ideal-result}, and the only varied
  parameter was $\sS$. In the selected geometries, the shapes and
  sizes of the grafts were such that with sufficiently high
  interface conductance $\sS$, we observed breakthrough of excitation into the
  host tissue (with a few exception of some geometries at some
  resolutions, as will be explained in due course).  We have used the
  bisection method to determine the threshold value of $\sS$ between
  propagation and no-propagation.

\Figure{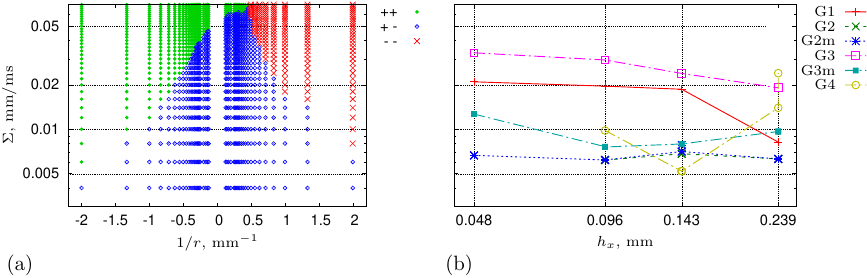}{!htb}{Break-through thresholds for the graft/host
  interface conductance.}{
  (a) The diagram of boundary curvature $1/\R$ vs interface
    conductance $\sS$, obtained from idealized geometry simulations:
    $1/\R>0$ means convex boundary as in \fig{RS-portrait-rev}, and
    $1/\R<0$ means concave boundary, so host tissue is within the
    quarter-disk of radius $\R$ and graft tissue is outside it. The
  meaning of colours and symbols is the same as in \fig{RS-portrait-rev};
  the penetration threshold corresponds to the lower edge of the area
  of filled green circles.
  (b) Penetration thresholds obtained for the realistic geometries at
  various resolutions $\ss$. The horizontal axis in (b) is
  logarithmic. The vertical $\sS$ axis is logarithmic and common for both panels. 
  }  

  As we shall see below, the analysis of this threshold in realistic
  geometries has revealed the importance of cases in which the
  graft/host boundary is not \emph{convex}, as in \fig{RS-portrait-rev}, where
  the host tissue ``surrounds'' the graft, but \emph{concave}, where the
  host tissue ``wedges into'' the graft. Motivated by these
  observations, we have repeated the simulations underlying
  \fig{RS-portrait-rev}, this time swapping the roles of the graft and
  host tissues, with the host inside the circle and the graft outside
  it, thereby reversing the sign of the boundary curvature relative to
  \fig{RS-portrait-rev}. The two series of simulations are combined into a
  single parametric portrait, shown in \fig{thr-graphs}(a), now
  plotting $\sS$ against curvature $1/\R$, rather than
  against $\R$ itself, so as to represent positive and
  negative curvatures more conveniently in the same diagram.

  The thresholds found by bisection method in the histology-derived
  geometries introduced above,
  \geom{CX122-3},
  \geom{CT110-3},
  \geom{CT115-3} and
  \geom{CT110-4},
  are shown in \fig{thr-graphs}(b). For two of these, \geom{CT110-3} and
  \geom{CT115-3}, we considered variations, called \geom{CZ110-3} and \geom{CC115-3}
  respectively, which shall be explained in due course below. The
  \geom{CT110-4} curve has two distinct values at $\ss=0.239\,\mm$; this is
  also discussed below. 
  For now
  we just note that in view of the range of the values of the
  thresholds and significance of their relative rather than absolute
  variations, we use the logarithmic scale for the threshold $\sS$,
  and in order for data represented in panels (a) and (b) immediately
  comparable, we make the $\sS$ axis in (a) logarithmic too, and make
  the vertical axis both diagrams identical. 

  \Figure{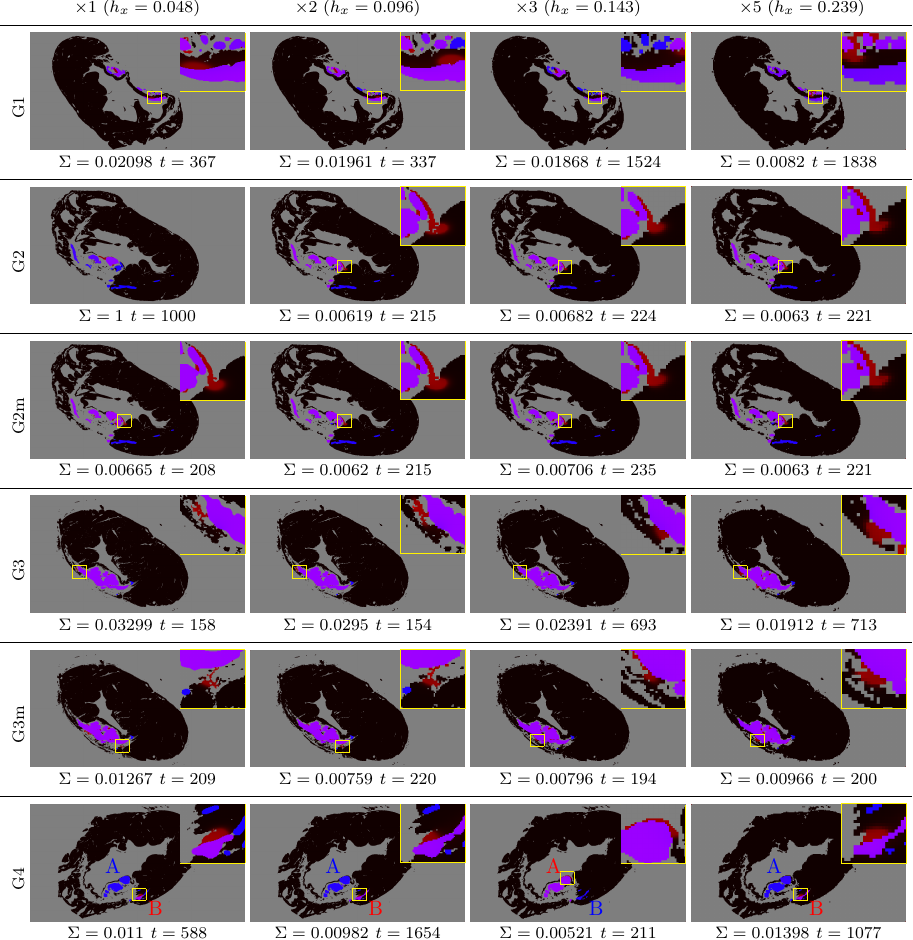}{!htb}{First break-throughs.}{%
    See details in the text.
      }

  A detailed analysis of the excitation breakthroughs in the
  individual geometries at various spatial resolutions is summarised
  in \fig{fb-all}.  Each row of panels in that figure corresponds to
  one geometry, as labelled on the left.  Each panel is a snapshot of
  simulation, with spatial resolution $\ss$ in $\mm$, as designated at the top,
  the interface conductance $\sS$ in $\mm/\ms$ and time moment $\t$
  in $\ms$ as indicated
  at the bottom. The value of $\sS$ for each panel is chosen for
  $\Delta\sS\in(10^{-5},2\cdot10^{-5})$ above the threshold found by
  bisection. The time moment was chosen as close as possible to the
  moment of breakthrough so that its location is evident.
  Most panels are supplied with insets showing interesting sites,
  typically where the breakthrough has
  happened, with $5\times$ magnification.
  Now we shall discuss each of the rows of \fig{fb-all} one
  by one starting from the top.

  For \geom{CX122-3}, the threshold is relatively large compared to other
  geometries, and is reasonably consistent across
  resolutions. Inspection of the penetration sites shows that they are
  different in different resolutions but all are in roughly the same
  area of the graft/host interface characterized by relatively small
  curvature, i.e. nearly straight. Our interpretation is in terms of
  the proverb ``the worst spoke in a cart breaks first'', and in the
  present case the ``spokes'' are of similar quality so which is worst
  depends on the perturbation caused by the spatial discretization.

  In contrast, \geom{CT110-3} demonstrates are 2- to 3-times lower threshold,
  rather consistent between resolutions \X2 through to
  \X5. The first break-through happens in the same place in
  these three resolutions, and this place is characterised by the thin
  strip of host tissue wedged between graft tissue and insulating
  surrounding. Somewhat simplifying, one can argue that the acute
  corner of this wedge corresponds to large negative local curvature
  of the graft/host interface, hence the lower threshold value of
  $\sS$. Significantly, the original (\X1)
  resolution is distinct in that the excitation never propagates from
  the graft into the host tissue. The magnified view of the usual
  penetration site shows that in the full resolution, there is in fact
  no contact between host and graft, and this contact appears only at
  increased space steps as a result of coarsegraining.

  We have directly verified this interpretation by manual edition of
  \geom{CT110-3} geometry, adding a few (to be precise, 228 pixels, about
  $0.014\%$ of the total) tissue nodes in the cleft that disconnected
  the piece of host tissue wedging into the graft and the bulk,
  producing a variation of the geometry we labelled as \geom{CZ110-3}. As can
  be observed both in \fig{thr-graphs}(b) and in the \geom{CZ110-3} row of
  \fig{fb-all}, this did not affect the threshold at lower resolutions,
  which was to be expected because actual penetration happened not at
  the edited site, but now the penetration happened, at very much the
  same threshold, also at the full \X1 resolution. Note that the threshold,
  wherever measured, is very consistent between the resolutions in
  \geom{CT110-3} and \geom{CZ110-3}, which we attribute to the fact that the shape
  of the critical piece of the interface is well defined and survives the
  coarsening.
  
  For \geom{CT115-3}, the first break-through site remains approximately the
  same in all resolutions, but the graft/host interface is rather
  rugged so its local geometry changes noticeably with resolution. Hence
  \X1 and \X2 resolutions concur both in timing (during the first
  oscillation of the graft) and the threshold
  of $\sS$, whereas \X3 and \X5 occur much later (on the second
  oscillation) and, at \X5, a somewhat smaller $\sS$: here we can
  see a host ``wedge'' similar to the one that ensured easy passage in
  \geom{CT110-3} and \geom{CZ110-3} we saw earlier.

  \geom{CC115-3} is a variation of \geom{CT115-3}, in that it is obtained from the
  same source histology image, with a single parameter of image
  processing modified: the intensity threshold used to identify host
  tissue changed by $5\%$ hence the host tissue area is somewhat
  wider. We observe that this gives a noticeable widening of the host
  tissue but most importantly, moves the graft/host interface,
  hence position, timing and threshold $\sS$ of first breaks changes
  completey compared to \geom{CT115-3}. Resolutions $\X1$ and $\X2$ in this
  geometry well agree, but $\X3$ and $\X5$ choose different
  sites. Since timing and threshold $\sS$ in $\X3$ and $\X5$
  are not too different from those in $\X1$ and $\X2$, we conclude
  that in this case the two sites are ``spokes'' of similar quality
  and which of them breaks first depends on minor detail depending on
  resolution, similar to the \geom{CX122-3} story.

  \Figure{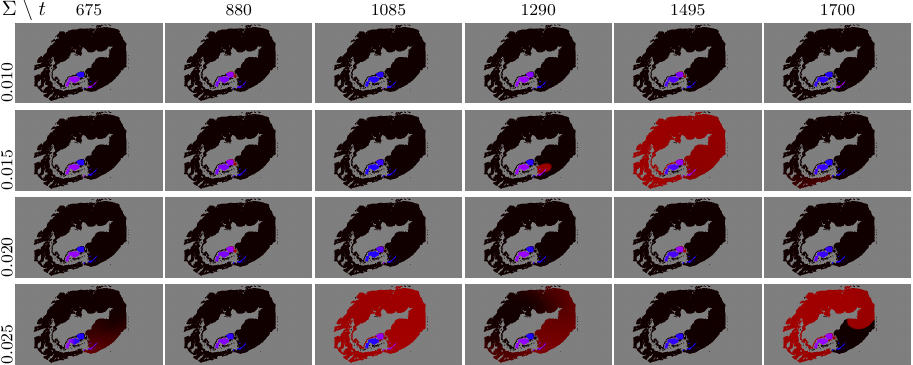}{!htp}{Four alternative scenarios in \geom{CT110-4}.}{
    Selected panels from selected simulations at \X5 resolution. 
    Each row corresponds to one selected value of $\sS$ as indicated
    on the left, in $\mm/\ms$. Each column corresponds to one selected value of $\t$
    as indicated on the top, in $\ms$. 
    }
  
  Finally, we are looking at \geom{CT110-4}, represented by the last row. The
  situation here is more interesting, as depending on the resolution
  and on $\sS$, the excitation propagation may occur from
  one of two alternative graft clusters, from the massive ``A'',
  connected to the bulk of the host tissue via a thin stripe, as seen
  in the \X3 panel, or from the slender ``B'', embedded deep into the
  bulk of the host tissue, as seen in \X1, \X2 and \X5 panels (note
  that threshold $\sS$ is reasonably consistent between the last three
  and distinct from \X3). The new feature here is that depending on
  $\sS$, the slender graft B may oscillate or may be
  completely suppressed by electrotonics of the surrounding host
  tissue, which is distinct from previous geometries considered, where
  the graft that caused excitation of the host was always oscillating
  and the question was only whether its excitation will or will not penetrate the
  interface.
  In \geom{CT110-4}, the situation is more complicated, and \emph{four}
  qualitatively different scenarios can be identified. Which of these
  occurs depends on the resolution, owing to resolution-dependent
  changes in the connections (a situation similar to that described
  above for \geom{CT110-3}). At the \X5 resolution, however, all four
  scenarios were observed, and we illustrate them in
  \fig{alternative}.
  Namely, for $\sS\in[0,0.013]$, both the massive
  graft A and the slender graft B oscillate, asynchronously, without
  exciting the host, as in the top row in \fig{alternative}. For
  $\sS\in[0.014,0.016]$, the interface conductance is still insufficient for
  excitation penetration from A, but B
  successfully ignites the host, as in the second row of
  \fig{alternative}. For $\sS\in[0.017,0.024]$, no excitation of the
  host tissue happens, because the conductance is too small for
  penetration from A, but is large enough to suppress
  oscillations of B so it remains quiescent
  throughout, as in the third row of \fig{alternative}. Finally, for
  $\sS\ge0.025$, while the own oscillations of B
  remain suppressed, the conductance is sufficient for excitation
  breakthrough from A, as in the bottom row of
  \fig{alternative}.  Note that the two different routes of excitation
  breakthrough, from B and from A,
  have different corresponding $\sS$ thresholds, which explains why
  there are two symbols at the same \X5 on the \geom{CT110-4} curve in
  \fig{thr-graphs}(b).

\section*{Discussion}

In this study, we formulated an inner-boundary mathematical (PDE)
description of the interaction between hPSC-CM graft and host tissues
and developed a method for approximating this description in
finite-difference computations. We validated the finite-difference
approximation by demonstrating numerical convergence in problems with
exact analytical answers. We also developed a finite-width boundary
zone imitation of the inner boundary that can be implemented using
existing finite difference (\BB) and finite element (\OC)
computational platforms. We demonstrated that this approximation
provides acceptable accuracy for many practical applications and
applied our framework to investigate graft-host interactions in both
idealized (circular) and anatomically realistic
(histology-derived) geometries. Our findings highlight the importance
of graft size and shape, interface conductance, and interface curvature in
determining whether spontaneous graft activity is suppressed, remains
confined to the graft, or propagates into surrounding host
myocardium. This suggests that the “critical mass” of a graft’s
arrhythmogenic potential is not solely a function of cell number, but
a delicate balance of local diffusivity and ionic current density.

From
a theoretical perspective,
it is quite possible that a stochastic description of the graft-host interface
becomes
equivalent to the inner boundary description in the limit of infinitely fine resolution.
However, there are
at least two
challenges
associated with
this assertion.
The first is
technical: how to relate the percentage of
severed
connections in the
stochastic description to the conductance of the inner boundary
description,
particularly since
this relationship will
vary with
different $\ss$. The other is more
fundamental
and
arises
from the
highly
nonlinear
nature
of the excitation propagation.
Consequently
even a small region of a well-connected interface may suffice
for a breakthrough.
Thus, as few as
$1\%$ surviving connections at
coarse
resolution may be functionally equivalent to full
connectivity,
whereas the $\ss\to0$ limit
may
require a
substantially greater
fraction of
surviving
connections.
As our examples illustrate, the gradual inner
boundary description
remains across resolutions.

  It is worth considering some of the potential implications of the
  methods described in this study to regenerative therapies based on
  hPSC-CM transplantation and their potential for electromechanical
  integration and engraftment arrhythmias. For example, one potential
  application of the method presented here is modelling heterotypic
  gap junctions between graft and host tissues that may arise from
  mismatched hemichannels (i.e. Cx40 or Cx45 in the graft versus Cx43
  in the host). Although heterotypic gap junctions are relatively
  uncommon compared to the homotypic counterparts, they may form in
  the cell transplantation context because connexins such as Cx40 and
  Cx45 may be expressed in off-target cardiomyocyte subtypes such as
  nodal cells\cite{37686171}, which have been identified in ventricular hPSC-CM
  differentiation protocols\cite{39196193}. Moreover, connexin-45 has also been
  identified in the heart during early developmental stages\cite{10381888,12619874},
  consistent with the immature phenotype of hPSC-CMs, and has been
  detected in embryonic stem cells-derived cardiomyocytes under stimulated
  conditions\cite{25080381}.
  Heterotypic gap junctions have different unitary conductances
  from homotypic gap junctions\cite{17189315}; this can be directly captured
  using the method developed here.

  Although the analysis presented here was limited to select
  histological sections, the application of our developed methods
  extends beyond this study. For example, it is possible that this
  approach can be retrospectively applied to large animal hPSC-CM
  transplantation studies to identify possible explanations for
  observed differences in arrhythmia burden between different animals,
  cell products, or doses. Pairing this with ECG monitoring data would
  also be useful to determine whether end-point graft morphology
  correlates with arrhythmia burden throughout the study. This may be
  particularly informative in understanding how injection site
  contributes to the arrhythmogenic potential of a cell product, given
  that the location of delivery determines the extent and geometry of
  host myocardium interacting with the graft. By screening graft
  morphologies and phenotypes through this computational pipeline, we
  can identify the specific graft volumes and interface conductance ranges
  required to ensure stable pacing, offering a robust methodology to
  mitigate the arrhythmia risks associated with cardiac cell
  therapies. Such analyses would benefit from further refinement by
  directly comparing optical mapping of cardiac slices with
  computational modelling of paired histological sections to provide
  key validation data.

  The physiological relevance and predictive utility of our model
  depend on how accurately its anatomical and electrophysiological
  assumptions reflect the properties of engrafted hPSC-CMs and their
  interactions with host myocardium. For instance, a more recent
  hPSC-CM kinetic model \cite{Gibbs-Boyle-2026} predicts substantially faster spontaneous
  oscillations. This is important because the maximum spontaneous
  firing rate in our models was determined by the underlying
  single-cell behaviour, while several groups have reported
  tachyarrhythmias with rates exceeding 150~bpm following hPSC-CM
  transplantation in large-animal models, with some cases reaching
  200~bpm \cite{31056479,37028405,39196193,34506727}.
  Therefore, additional modifications of the ionic
  model may be necessary to reproduce the higher rates observed in
  vivo. Beyond spontaneous firing rate, current models assume
  homogeneous hPSC-CM electrophysiological properties within
  grafts. However, this assumption may not reflect the cellular
  heterogeneity reported in hPSC-CMs both before and after
  transplantation\cite{31056479,37028405,39196193,34506727}.
  Heterogeneity in action potential
  morphology may further influence impulse propagation both within the
  graft and across the graft-host interface. Finally, graft
  cardiomyocytes may also interact with host Purkinje fibres\cite{39196193}. Such
  interactions may differ from the graft-host interactions modelled
  here, potentially altering whether a given graft can initiate an
  impulse that will initiate excitation in the surrounding host
  myocardium.

  One limitation of the proposed approach here is the difficulty of
  directly measuring the interface conductance $\sS$ between graft and host
  tissues, a challenge comparable to that of directly measuring the
  voltage diffusivities in either tissue compartment. Indeed,
  simulations commonly rely on voltage diffusivities inferred from
  indirect measurements, such as conduction velocity, rather than
  direct physical measurements of intercellular resistivities, cell
  membrane capacitance, and surface/volume ratios, see for example
  \cite[pp.~H1090--H1091]{TenTusscher-Panfilov-2006}
  and
  \cite[p.~6]{openCARP-paper}.
  This limitation does not
  invalidate such simulations, but does warrant caution in
  interpreting their results. The mathematical form of the
  description may be more important than the precise values of the
  diffusivity tensor components. Similarly, we propose that our
  approach may remain useful even when $\sS$ cannot be directly measured
  and must instead be inferred from indirect evidence, provided that
  the results are interpreted with appropriate caution.

\section*{Conclusion}

The host-graft interface acts as a dynamic impedance barrier,
where spontaneous oscillations of the graft and the propagation of
excitation into the host tissue are governed by a strict source-sink
relationship that depends on the inner boundary's conductance and
curvature.
Our results are subject to obvious
  limitations, including choice of the models of cellular excitability,
  choice of histology images, choice of the many parameters of the
  conversion of the images into computational grids, and
  number of simulations performed. Notwithstanding these limitations,
  we think that some generic conclusions could be made, if not
  definitive, then at least hypothetical, as suggestions for further,
  more detailed research. 
In particular:
\begin{itemize}
\item Spontaneous oscillations in sufficiently small grafts can be
  suppressed by the electrical load of the host tissue. The critical
  graft size is reduced as the interface conductance decreases.
\item
  When spontaneous oscillations in the graft are not suppressed,
  they may stay isolated within the graft if interface conductance
  is small, or engage the host tissue if the interface conductance
  is large.
  The critical conductance that distinguishes isolation vs
  engagement depends on the curvature of the interface. A convex
  interface, when the smaller mass graft tissue is surrounded by
  a larger mass of host tissue, prevents engagement; a concave
  interface, when the larger graft mass surrounds a smaller host mass,
  facilitates engagement.
\item Graft oscillations slow down considerably before being
  suppressed, when convex interface curvature increases at the same
  interface conductance, and slow down slightly if the interface
  conductance increases at the same curvature, with a slight jump
  of in the oscillation period between isolated oscillations and
  host engagement.
\item In our ``realistic'' geometries we observed considerable
  decrease in the critical conductance due to the concave
  interface. This indicates potential important role of seemingly
  insignificant features of the graft/host interface, as, at least
  in our models, breakthrough from a very small,
  $\sim1\,\mm$, piece of boundary, may be a potentially
  arrhythmogenic event.
\item In our cellular kinetics models, a decrease of the frequency of the graft
  spontaneous oscillations at smaller graft sizes was more than
  two-fold, which again may be significant for arrhythmogenicity, as
  it suggests that small graft islets may be
  more easily
  entrained by
  the host's
  sinus
  rhythm even if not suppressed entirely by the host electrical
  load, while large graft sizes may oscillate faster than
  the sinus rate
  and
  thus become ectopic foci.
\end{itemize}
This suggests that the ``critical mass'' of a graft is not solely a
function of cell number, but a delicate balance of local diffusivity
and ionic current density. By ``pre-screening'' graft phenotypes
through this computational pipeline, we can identify the specific
graft volumes and interface conductance ranges required to ensure stable
pacing, offering a robust methodology to mitigate the risks
associated with cardiac regenerative therapies.

\section*{Supporting information}

\paragraph*{S1 Appendix.}
\label{S1_Appendix}
{\bf Details and tests of numerical approximations.}
Approximation of non-flux boundary conditions; approximation of
inhomogeneous anisotropic diffusion and its test; suppression of
numerical instability in strong anisotropy.

\section*{Acknowledgments}

We acknowledge the support of the Government of Canada’s New Frontiers in Research Fund (NFRF), [NFRFT-2022-00447] and Canada Research Chairs Program (CRC-2020-00245). 

\ifpreprint\else\nolinenumbers\fi



\clearpage

\appendix

\section{S1 Appendix. Details and tests of numerical approximations}

\subsection{Approximation of non-flux boundary conditions}
\seclabel{nonflux-appr}

When discussing finite-difference approximation, we shall be using the
language of electrical circuits: each node corresponds to an element
with the capacitance together with nonlinear active resistances
representing the local processes described by $\f()$ and $\G()$ in
\eq{monodomain},  whereas a finite-difference approximation of
diffusion operator $\nabla\Dhat\nabla$  corresponds to the electrical
connection between the nearest nodes, with the number of neighbours
defined by the stencil, that is 4 neighbours for a 5-point stencil and 8
neighbours for a 9-point stencil, and the weights in the stencils
correspond to the conductances of the connections.
In this language, our \BB implementation of 
the non-flux boundary conditions
amounts to removing the connection between the nodes if
one node belongs to the tissue and the other does not. This
is a popular finite-difference approach, described
e.g. in~\cite{TenTusscher-Panfilov-2008}, and is equivalent to the
standard non-flux boundary conditions approximation in \OC, to the
extent finite differences can be compared to finite elements. 
As reported in \cite{Antonioletti-etal-2017}, this gives
approximation of the order roughly $\O{\ss^{3/2}}$, i.e. worse than
the $\O{\ss^2}$ approximation of the diffusion operator itself, but acceptable in
practice.

In assessing this approximation and the approximations discussed
below, one has to bear in mind one principal consideration. In an
abstract mathematical setting, it is typically assumed that PDE is set
in a given domain, and refining the spatial discretization yields increasingly 
accurate approximations of that domain. A good approximation
of boundary conditions in a domain of a complicated ``arbitrary''
shape on a rectangular grid should consider the position of
the boundary with respect to grid nodes, as discussed e.g. in
\cite[\S20.9 pp.198--202]{Forsythe-Wasow-1960}
and
\cite[\S4.1.4 pp.248--255]{Samarskii-2001}
for the case of Dirichlet
boundary conditions. Putting aside the fact that this is a rather
complicated procedure, and is even more complicated for the non-flux
boundary conditions we need, see e.g. a brief discussion in
\cite[\S20.10 pp.202--204]{Forsythe-Wasow-1960},
the fundamental fact is that,
when the domain is obtained from experiment, say MRI or microscopic
data, there is no ideal ``exact domain'' to be approximated by the grid; on
the contrary, the domain is defined only in terms of the grid. That
is, the question of exactly where lies the boundary between the last
tissue point and the nearest void point is purely ``academic'', as
this information is simply unavailable in practice. Meanwhile, the
displacement of the (imagined) boundary between the grid points means
a local error of the solution of the order of $\O{\ss}$.

\subsection{Approximation of inhomogeneous anisotropic diffusion}
\seclabel{av-appr}

A popular approach to discretization of the spatially-nonhomogeneous
anisotropic diffusion operator is, as described e.g. in
\cite{Clayton-Panfilov-2008}, using Einstein's index summation
convention,
\[
  \nabla\left( \Dhat \nabla \V \right)
  =
  \df{}{\x_\i} \left( \D_{\i\j} \df{\V}{\x_\j} \right)
  =
  \C_\j \df{\V}{\x_\j}  + \D_{\i\j} \dff{\V}{\x_\i}{\x_\j},
  \qquad
  \C_\j = \df{\D_{\i\j}}{\x_\i},
\]
with central difference approximation of spatial derivatives of $\V$
and $\D_{\i\j}$; in \BB, that the diffusivity tensor is defined on the
same spatial grid as the diffusive field $\V$, but one can find more
sophisticated approaches using e.g. a staggered grid for $\D_{\i\j}$.
This scheme is not conservative but that is of little consequence for
most cardiac excitability computations where the spatial dependence of
the diffusivity tensor is smooth, since the transmembrane currents
violate conservation of ``diffusing'' electrical charge anyway.
However, in the present study we need to look at the \emph{discontinuous}
diffusivity tensor, and in such cases, the above non-conservative
scheme does not converge, see e.g. again the classical textbook
\cite[\S3.2.1 p. 147--149]{Samarskii-2001}. The general idea of
constructing a conservative scheme is given in the same book, see
e.g. \cite[p. 156]{Samarskii-2001} for the formula for 1D diffusion
with diffusivity defined on the same grid as the diffusive field. This
approach extends to 2D and 3D, but the resulting formulas are not
easy to find
in the literature, so we present the 2D scheme used in our
computations here.
In the following formulas we let $\x_1=\x$, $\x_2=\y$.
We assume the diffusivity tensor $\D_{\i\j}(\x,\y)$ is defined on the
same square grid as the diffusing field $\V(\x,\y)$. To introduce
notations, let the $(\x,\y)$ grid be $\x_\k\appr \x_0+\k\ss$,
$\y_\l\appr \y_0+\l\ss$, $\k,\l\in\Zahlen$,
$\V_{\k\l}\appr \V(\x_\k,\y_\l)$,
$\D_{\i\j}^{\k\l}\appr \D_{\i\j}(\x_\k,\y_\l)$, where $\appr$ denotes
the relationship between an object related to the PDE and its
discretization counterpart. The conservative scheme derived using
the ``integro-interpolation method'' in \cite[p.~156]{Samarskii-2001} for
the 1D case, $\D_{11}\equiv\D$, says
\[
  \@_\x \bigg( \D(\x) \, \@_\x \V(\x) \bigg)
  \appr
  \frac{1}{\ss} \left(
    \D^{\k+1/2} \, \frac{\x_{\k+1}-\x_\k}{\ss}
    -
    \D^{\k-1/2} \, \frac{\x_{\k}-\x_{\k-1}}{\ss}
  \right)
\]
where the half-step diffusivity values are defined as mean values
between those at the centre point of the stencil and the neighbouring
value,
\[
  \D^{\k+1/2} = \mean{\D^\k}{\D^{\k+1}},
  \qquad
  \D^{\k-1/2} = \mean{\D^\k}{\D^{\k-1}},
\]
and the mean could be understood as arithmetic, 
\[
  \mean{\a}{\b} \bydef \frac{\a+\b}{2},
\]
(averaging the conductivities)
or harmonic,
\[
  \mean{\a}{\b} \bydef \frac{1}{(1/\a+1/\b)/2} = \frac{2 \a \b}{\a + \b}
\]
(averaging the resistivities). 
Both variants can be found in literature; the choice between the two
is ``academic'' for the same reasons as discussed in \secn{nonflux-appr} above.
We found little difference
between the two in our calculations. This can be extended to 2D and 3D. For our purposes,
the 2D scheme works out thus:
\[ 
  \begin{array}[t]{ccccccc}
    \nabla\left( \Dhat \nabla \V \right)
    & \appr & \w_{\m\p}\V_{\m\p} &+& \w_{\o\p}\V_{\o\p} &+& \w_{\p\p}\V_{\p\p} \\
    &       + & \w_{\m\o}\V_{\m\o} &+& \w_{\o\o}\V_{\o\o} &+& \w_{\p\o}\V_{\p\o}  \\
    &       + & \w_{\m\m}\V_{\m\m} &+& \w_{\o\m}\V_{\o\m} &+& \w_{\p\m}\V_{\p\m}  ,
  \end{array}
\]
where the subscripts are abbreviated to represent shifts from the
centre of the stencil, e.g. $\V_{\m\p} \bydef \V_{\k-1,\l+1}$,
$\V_{\o\m}\bydef \V_{\k,\l-1}$ etc.,  and the stencil weights are defined
as
\begin{align}\eqlabel{ani-stencil}
  \begin{array}[t]{cccccc}
    \w_{\m\o} &=\frac{1}{\ss^2} \mean{\D_{11}^{\k,\l}}{\D_{11}^{\k-1,\l}}, & \w_{\p\o} &=\frac{1}{\ss^2} \mean{\D_{11}^{\k,\l}}{\D_{11}^{\k+1,\l}}, \\
    \w_{\o\m} &=\frac{1}{\ss^2} \mean{\D_{22}^{\k,\l}}{\D_{22}^{\k,\l-1}}, & \w_{\o\p} &=\frac{1}{\ss^2} \mean{\D_{22}^{\k,\l}}{\D_{22}^{\k,\l+1}}, \\
    \w_{\m\m} &=\frac{1}{2\ss^2} \mean{\D_{12}^{\k,\l}}{\D_{12}^{\k-1,\l-1}}, &
    \w_{\p\m} &=-\frac{1}{2\ss^2} \mean{\D_{12}^{\k,\l}}{\D_{12}^{\k+1,\l-1}}, \\
    \w_{\m\p} &=-\frac{1}{2\ss^2} \mean{\D_{12}^{\k,\l}}{\D_{12}^{\k-1,\l+1}},
    &
    \w_{\p\p} &=\frac{1}{2\ss^2} \mean{\D_{12}^{\k,\l}}{\D_{12}^{\k+1,\l+1}} , \\
    \w_{\o\o} &= - \sum\limits_{ \substack{ p,q\in\{\m,\o,\p\}
        \\
        (pq)\ne(\o\o) } } \w_{pq} .&&&
  \end{array}
\end{align}
The conservative property of this scheme follows directly from the symmetry of the
function
$\mean{\D^\anode}{\D^\bnode}\equiv \mean{\D^\bnode}{\D^\anode}$,
which implies that if nodes $\anode$ and $\bnode$ are connected,
then the current received by $\anode$ from $\bnode$ is
$\I_{\bnode\anode}=\left(\V_\anode-\V_\bnode\right) \mean{\D^\anode}{\D^\bnode}$,
and is exactly opposite to that
received by $\bnode$ from $\anode$, 
$\I_{\anode\bnode}=\left(\V_\bnode-\V_\anode\right) \mean{\D^\bnode}{\D^\anode}$,
so $\I_{\bnode\anode}+\I_{\anode\bnode}=0$ and
total charge in the system is conserved.

The
approximation property is verified in the standard way using Taylor
expansions; these calculations are straightforward but bulky and are
omitted here.

Note that in the isotropic case, $\D_{12}\equiv0$, Eq.~\eq{ani-stencil} automatically
gives $\w_{\pm,\pm}=0$, that is the stencil reduces from 9-point to
5-point.

\subsection{Test of inhomogeneous anisotropic diffusion approximation}
\seclabel{av-test}

We tested the approximation property of the scheme described in \subsecn{av-appr}
on an 
artificially constructed example with an exact analytical
solution. This was a reaction-diffusion equation, in which the
diffusivity is anisotropic and space-dependent, and the reaction term
is linear and constructed in such a way as to ensure that a given
formula provides an exact solution of the PDE.
Specifically, in terms of polar coordinates $(\polrad,\polang)$,
$\x=\polrad\cos\polang$, 
$\y=\polrad\sin\polang$, 
we considered the annular
domain, $\Domain=\{(\x,\y) \,|\,  \polrad \in(\Ri,\Ro)\}$, the
fibres in the radial direction, 
and the PDE
\[
  \df{\V}{t} = \nabla\left( \Dhat \nabla \V \right)
  +
  \frac{\aveps}{\polrad} \cos(\avm\polang) \,\ee{-\avlam\t}  \big( \left(1+\avk^2\polrad^2\right)\Bes1(\avk \polrad) - \avk\polrad\Bes0(\avk\polrad) \big),
\]
where
\[
  \Dhat=\Mx{\D_{\i\j}}, \quad
  \D_{11} = \DT + (\DL-\DT)\cos^2\polang, \quad
  \D_{12}=\D_{21}=(\DL-\DT)\cos\polang\sin\polang,  \quad
\]\[
  \D_{22}= \DT + (\DL-\DT)\sin^2\polang, \quad
  \DL(\x,\y)=1+\aveps\polrad, \quad \DT(\x,\y)=\DL(\x,\y))/\avm^2, 
\]
$\avk=\besroot2$,
$\Ro=1$, $\Ri=\besroot1/\besroot2$,
$\avlam=\Ri^2$,
$\aveps=1/2$,
$\Bes{\besind}()$ is Bessel function of first kind of index $\besind$
and
$\besroot\besind$ is $\besind$-th positive root of $\Bes1'$,
i.e. $\besroot1=1.84\dots$, $\besroot2=5.33\dots$,
all quantities are dimensionless. 
This problem has the exact solution
\begin{equation}\eqlabel{avsol}
  \V(\x,\y,\t) = \Bes1(\avk\polrad)\cos(\avm\polang)\,\ee{-\avlam\t} .
\end{equation}

\subsection{Tackling numerical instability in strong anisotropy}
\seclabel{polishing}

Finite-difference approximation of the anisotropic diffusion operator
is known to potentially cause numerical instability by failing to
produce positively definite discretization, see %
e.g.~\cite[\S3.4.2 pp.88--95]{Weickert-1998}. %
This can be easily seen from the stencil of the discretization of the
anisotropic diffusion \eq{ani-stencil}: for $\D_{12}>0$ we have
$\w_{\p\m}<0$ and $\w_{\m\p}<0$, and for $\D_{12}<0$, we have
$\w_{\m\m}<0$ and $\w_{\p\p}<0$.
  In terms of electric circuits, this
means nodes connected diagonally by negative resistivity, which is
inherently unstable.  This instability may be damped by surrounding
positive resistivities, if the anisotropy is not too high: according
to \cite[p.94]{Weickert-1998}, the critical value is
$\DL/\DT=3+2\sqrt2\approx5.83$, whereas our computations are only for
$\DL/\DT=0.255/0.0775\approx3.29$, so this damping works well for domains
with smooth boundaries.  However, ragged boundaries can lead to
undamped instability even at relatively mild anisotropy. The simplest
example is a ``dangling'', near-isolated point in the tissue, which
is connected to the bulk of the tissue by a single diagonal connection
with a negative weight. With the non-flux boundary conditions
discussed above, this means immediate instability related to this
``dangling'' point.  This difficulty may be addressed in various ways,
such as ``extremum preserving schemes'' \cite{Gao-Wu-2013-JCP},
``non-negative directional splitting'' enlarging the stencil until
positivity conditions hold \cite{Ngo-Huang-2016-CCP}, or mesh
refinement near the boundaries~\cite{Vogl-etal-2023-CMA} to cite some
recent ones.

For our finite-difference calculations, the anisotropy was strong
enough to cause problems, but not so severe as to necessitate the
complicated approaches referenced above. We adopted a simpler
practical approach: we modified the computational grid, not by
refining it, but by smoothing the boundaries by eliminating nodes
which de-facto cause instabilities. Namely, we utilized the following
algorithm:

\begin{algorithm}[H] 
  \caption{``Grid polishing''}
  \begin{algorithmic}[1]
    \Require $\Amax>1$, $\Afrac\in(1/\Amax,1)$, $\Tmax>0$. 
    \State Start with the given tissue points on the grid, including both host and
    graft (henceforth ``tissue'') with anisotropic diffusion
    tensor as found via the Laplace equation in host and isotropic
    diffusion in graft.  %
    \Repeat %
    \State Remove the ``absolutely unstable points'' with $\w_{\o\o}>0$ from the tissue grid. 
    \State Assign random
    values uniformly distributed in $[-1,1]$ to the tissue points, as the initial
    condition.  %
    \Repeat\ Solve anisotropic diffusion equation on the tissue
    with non-flux boundary conditions, starting from this initial condition. %
    \Until\ Maximum absolute value of the solution exceeds $\Amax$ or the total model
    time exceeds $\Tmax$.  %
    \State Identify all tissue points where the absolute
    value of the solution exceeds $\Afrac\Amax$ and remove them from the
    tissue.  %
    \Until\ the run time in the last initial-value problem exceeded
    $\Tmax$.  %
    \State \Return the remaining tissue points as the new
    ``polished'' tissue grid, retaining their original host/graft labelling
    and anisotropic diffusion tensor.
  \end{algorithmic}
  \label{alg:polishing}
\end{algorithm}

The idea of this algorithm is that any true solution of the anisotropic diffusion
equation with non-flux boundary condition will tend with time to a constant,
and with a given initial condition this constant must be within
$[-1,1]$. Hence at $\Amax>1$, the algorithm will remove any points only if
instability is present, and the expectation is that if $\Amax$ is large enough,
this value is reached first at the nodes which have \emph{caused} that
instability (in this aspect, the points with $\w_{\o\o}>0$ are considered
  unstable \textit{a priori}, so their elimination just speeds up the iterations).
This argument is of course only heuristic, but it worked well for
our grids. We used $\Amax=3$, $\Afrac=0.8$, $\Tmax=10^3$; the diffusion
equation  was with  diffusivities $\DL=1$,
$\DT=0.303922$,
i.e. the same anisotropy as used in actual simulations,
solved using explicit Euler time stepping with time step $\ts=0.1$
and space step $\ss=1$, all quantities dimensionless.

\end{document}